\documentclass[preprint,aps]{revtex4}
\usepackage{epsfig}
\newcommand{\be}{\begin{equation}}
\newcommand{\ee}{\end{equation}}
\newcommand{\ba}{\begin{eqnarray}}
\newcommand{\ea}{\end{eqnarray}}

\begin{document}

\draft

\title{Can the Brans-Dicke Gravity with $\Lambda $Possibly  be \\
a Theory of Dark Matter?}

\author{Hongsu Kim\footnote{e-mail: hongsu@kasi.re.kr}}

\address{International Center for Astrophysics, \\
Korea Astronomy and Space Science Institute, Daejeon 305-348,
Korea}


\begin{abstract}
The pure Brans-Dicke (BD) gravity with or without the cosmological
constant $\Lambda $ has been taken as a model theory for the dark
matter. Indeed, there has been a consensus that unless one
modifies either the standard theory of gravity, namely, general
relativity, or the standard model for particle physics, or both,
one can never achieve a satisfying understanding of the phenomena
associated with dark matter and dark energy. Along this line, our
dark matter model in this work can be thought of as an attempt to
modify the gravity side alone in the simplest fashion to achieve
the goal. Among others, it is demonstrated that our model theory
can successfully predict the emergence of dark matter halo-like
configuration in terms of a self-gravitating spacetime solution to
the BD field equations and reproduce the flattened rotation curve
in this dark halo-like object in terms of the non-trivial energy
density of the BD scalar field, which was absent in the context of
general relativity where Newton's constant is strictly a
``constant'' having no dynamics. Our model theory, however, is not
entirely without flaw, such as the prediction of relativistic jets
in all types of galaxies which actually is not the case.

\end{abstract}

\pacs{PACS numbers: 04.50.+h, 95.35.+d, 98.62.Gq, 98.80.Cq}

\maketitle

\narrowtext

\newpage

\section{Introduction}

There has been a consensus among researchers that unless one
modifies either the standard theory of gravity, namely, general
relativity, or standard particle physics theory (say, the
Weinberg-Glashow-Salam's standard model), or both, one can never
achieve a satisfying understanding of the phenomena associated
with dark matter and dark energy. Along this line, in the present
work, we would like to propose an attempt to modify the gravity
side alone in the simplest fashion to achieve the goal. To be more
specific, the pure Brans-Dicke (BD) gravity \cite{bd} with or
without the cosmological constant $\Lambda $ shall be taken as a
model theory for the dark matter. Indeed, the BD theory is the
most studied, and hence the best-known, of all the alternative
theories of classical gravity to Einstein's general relativity
\cite{will}. This theory can be thought of as a minimal extension
of general relativity designed to properly accomodate both Mach's
principle \cite{will, weinberg} and Dirac's large number
hypothesis \cite{will, weinberg}. Namely, the theory employs the
viewpoint in which Newton's constant $G$ is allowed to vary with
space and time and can be written in terms of a scalar (``BD
scalar'') field as $G = 1/\Phi$. Besides, the BD scalar field (and
the BD theory itself) is not of quantum origin. Rather, it is
classical in nature and hence can be expected to serve as a very
relevant candidate to play some role in the late-time evolution of
the universe, such as the present epoch.

As a scalar-tensor theory of gravity, the BD theory involves an
adjustable, but undetermined, ``BD-parameter'' $\omega$, and as is
well-known, the larger the value of $\omega$, the more dominant
the tensor (curvature) degree and the smaller the value of
$\omega$, the larger the effect of the BD scalar. Also as long as
we select a sufficiently large value of $\omega$, the predictions
of the theory agree perfectly with all the
observations/experiments to date \cite{will}. For this reason, the
BD theory has remained a viable theory of classical gravity.
However, no particularly overriding reason has ever emerged to
take it seriously over general relativity. As shall be presented
shortly in this work, here we emphasize that it is the expected
existence of dark matter (and dark energy as well, see Ref.4) that
may put the BD theory over general relativity as a more relevant
theory of classical gravity consistent with observations that have
so far been unexplained within the context of general relativity.

As mentioned above, in our model theory for dark matter, the
effect of the cosmological constant $\Lambda $ shall be generally
considered. Here, $\Lambda $ is essentially supposed to play the
role of dark energy in the context of the BD theory as has been
studied in detail in Ref.4. As is well-known, the mysterious {\it
flattened} rotation curves observed for so long in the outer
regions of galactic halos have been the primary cause that called
for the existence of dark matter. Among others, therefore, we
shall demonstrate in this work that our model theory can
successfully predict the emergence of a dark matter halo-like
configuration in terms of a self-gravitating static and nearly
spherically-symmetric spacetime solution to the BD field equations
and reproduce the flattened rotation curve in the outer region of
this dark halo-like object in terms of the non-trivial energy
density of the BD scalar field, which is absent
in the context of general relativity where Newton's constant is strictly a ``constant'' having no dynamics. \\

\section{Schwarzschild-de Sitter-type solution in the BD theory of gravity}

As stated earlier in the introduction, we would like to
demonstrate in the present work that the BD gravity with or
without the cosmological constant can reproduce some
representative features of dark matter, such as the formation of a
dark matter halo inside of which the flattened rotation curves are
observed. Since the galactic dark matter halos are roughly static
and spherically-symmetric, we should, among others, look for such
dark matter halo-like solution to the BD field equations. Thus, in
this section, we shall construct the Schwarzschild-de Sitter-type
solution in the
BD theory and claim later on that it can represent the dark matter halo in the context of our model of dark matter. \\

As is well-known, even in Einstein's general relativity, Finding
exact solutions to the highly non-linear Einstein field equations
is a formidable task. For this reason, algorithms generating
exact, new solutions from the known solutions of simpler
situations have been actively looked for, and actually quite a few
have been found. In the BD theory of gravity, the field equations
are even more complex; thus, it is natural to seek similar
algorithms generating exact solutions from the already known
simpler solutions either of the BD theory or of the conventional
Einstein gravity. In particular, Tiwari and Nayak \cite{tina}
proposed an algorithm that allows stationary axisymmetric
solutions in the vacuum BD theory to be generated from the known
Kerr solution in the vacuum Einstein theory. Thus, in the present
work, we shall take the algorithm suggested by Tiwari and Nayak
\cite{tina} or by Singh and Rai \cite{sinrai} to construct the
Schwarzchild-de Sitter-type solution in the BD theory in the
presence of the cosmological constant from
the well-known Schwarzschild-de Sitter solution in Einstein gravity with the cosmological constant.  \\

Consider the BD theory in the presence of the (positive) cosmological constant $\Lambda $ described by the action
\begin{eqnarray}
S = \int d^4x \sqrt{g}\left[{1\over 16\pi}\left(\Phi R - \omega
{{\nabla_{\alpha}\Phi \nabla^{\alpha}\Phi }\over \Phi}\right) -
\Lambda \right],
\end{eqnarray}
where $\Phi $ is the BD scalar field representing roughly the
inverse of Newton's constant and $\omega $ is a generic parameter
of the theory. Extremizing this action with respect to the metric
$g_{\mu \nu}$ and the BD scalar field $\Phi $ yields the classical
field equations given, respectively, by
\begin{eqnarray}
G_{\mu \nu} &=& R_{\mu \nu} - {1\over 2}g_{\mu \nu}R + \frac{8\pi}{\Phi}\Lambda g_{\mu\nu} = 8\pi T^{BD}_{\mu \nu},
~~~\nabla_{\alpha}\nabla^{\alpha}\Phi = -{32\pi \over (2\omega + 3)}\Lambda ~~~{\rm where}  \nonumber \\
T^{BD}_{\mu \nu} &=& {1\over 8\pi}\left[{\omega \over \Phi^2}(\nabla_{\mu}\Phi \nabla_{\nu}\Phi
- {1\over 2}g_{\mu \nu}\nabla_{\alpha}\Phi \nabla^{\alpha}\Phi) + {1\over \Phi}(\nabla_{\mu}
\nabla_{\nu}\Phi - g_{\mu \nu}\nabla_{\alpha}\nabla^{\alpha}\Phi)\right].
\end{eqnarray}
The Einstein gravity is the $\omega \rightarrow \infty$ limit of
this BD theory. Now the algorithm of Tiwari and Nayak \cite{tina}
or Singh and Rai \cite{sinrai} goes as follows; In general, let
the metric for a stationary axisymmetric solution to the Einstein
field equations take the form
\begin{eqnarray}
ds^2 = - e^{2U_{E}}(dt + W_{E}d\phi)^2 + e^{2(k_{E}-U_{E})}[(dx^1)^2 + (dx^2)^2] +
h^2_{E}e^{-2U_{E}}d\phi^2
\end{eqnarray}
while letting the metric for a stationary axisymmetric solution to
the BD field equations be
\begin{eqnarray}
ds^2 = - e^{2U_{BD}}(dt + W_{BD}d\phi)^2 + e^{2(k_{BD}-U_{BD})}[(dx^1)^2 + (dx^2)^2] +
h^2_{BD}e^{-2U_{BD}}d\phi^2
\end{eqnarray}
where $U$, $W$, $k$, and $h$ are functions of $x^1$ and $x^2$
only. The significance of the choice of the metric in this form
has been thoroughly discussed by Matzner and Misner \cite{mm} and
Misra and Pandey \cite{mispan}. Tiwari and Nayak or Singh and Rai
first wrote down the Einstein and the BD field equations for the
choice of metrics in Eqs. (3) and (4), respectively. Comparing the
two sets of field equations closely, then, they realized that
stationary axisymmetric solutions of the BD field equations are
obtainable from those of the Einstein field equations provided
certain relations between metric functions hold. That is, if
$(W_{E}, ~k_{E}, ~U_{E}, ~h_{E})$ form a stationary axisymmetric
solution to the Einstein field equations for the metric in Eq.
(3), then a corresponding stationary axisymmetric solution to the
BD field equations for the metric in Eq. (4) is given by $(W_{BD},
~k_{BD}, ~U_{BD}, ~h_{BD}, ~\Phi )$, where
\begin{eqnarray}
W_{BD} &=& W_{E}, ~~~k_{BD} = k_{E}, ~~~U_{BD} = U_{E} - {1\over 2}\log \Phi, \\
h_{BD} &=& [h_{E}]^{(2\omega - 1)/(2\omega + 3)}, ~~~\Phi = [h_{E}]^{4/(2\omega + 3)}. \nonumber
\end{eqnarray}

Now what remains is to apply this method to obtain the
Schwarzschild-de Sitter-type solution in the BD theory from the
known Schwarzschild-de Sitter solution in Einstein gravity. To do
so, one needs some preparation, which involves casting the
Schwarzschild-de Sitter solution given in the usual Schwarzschild
coordinates $(t, r, \theta, \phi)$ in the metric form given in Eq.
(3) by performing the coordinate transformation (of $r$ alone)
suggested by Misra and Pandey \cite{mispan}. Namely, we start with
\begin{eqnarray}
ds^2 &=& - \left(1 - \frac{2M}{r} - \frac{8\pi
\Lambda}{3}r^2\right)dt^2 + \left(1 - \frac{2M}{r} - \frac{8\pi
\Lambda}{3}r^2\right)^{-1}dr^2 + r^2\left[d\theta^2 + \sin^2
\theta d\phi^2\right]
\nonumber \\
&=& - \frac{\Delta}{r^2}dt^2 + r^2 \left[\frac{dr^2}{\Delta} +
d\theta^2 + \sin^2 \theta d\phi^2\right],
\end{eqnarray}
where $\Delta = r^2(1 - 8\pi \Lambda r^{2}/3) -2Mr$ with $M$ and
$\Lambda$ denoting the ADM mass and the (positive) cosmological
constant, respectively. Consider now such a transformation of the
radial coordinate as the one introduced by Misra and Pandey
\cite{mispan};
\begin{eqnarray}
r = r(R) ~~~{\rm such ~that} ~~~\frac{dr^2}{\Delta} = dR^2.
\end{eqnarray}
Then, the Schwarzschild-de Sitter solution can now be cast in the
form in Eq. (3), i.e.,
\begin{eqnarray}
ds^2 = - \frac{\Delta}{L^2}dt^2 + L^2 \left[dR^2 + d\theta^2 \right] + L^2 \sin^2 \theta d\phi^2
\end{eqnarray}
with $L \equiv r(R)$ and, hence, $\Delta = L^2(1 - 8\pi \Lambda
L^{2}/3) -2ML$. Now, we can read off the metric components as
\begin{eqnarray}
e^{2U_{E}} = \frac{\Delta}{L^2}, ~~~W_{E} = 0, ~~~e^{2k_{E}} = \Delta, ~~~h^2_{E} = \Delta \sin^2 \theta.
\end{eqnarray}
Then, using the rule in Eq. (5) in the algorithm by Tiwari and
Nayak or by Singh and Rai, one finds that the metric components of
the Schwarzschild-de Sitter-type solution in the BD theory come
out to be
\begin{eqnarray}
e^{2U_{BD}} &=& \frac{\Delta}{L^2}\left(\Delta \sin^{2}\theta \right)^{-2/(2\omega+3)}, ~~~W_{BD} = 0, ~~~e^{2k_{BD}} = \Delta,  \\
h^2_{BD} &=& \left(\Delta \sin^{2}\theta \right)^{(2\omega -1)/(2\omega +3)},
~~~\Phi (R, \theta) = \left(\Delta \sin^{2}\theta \right)^{2/(2\omega+3)}.  \nonumber
\end{eqnarray}

Next, looking at the forms of the metric for a stationary
axisymmetric solution to the Einstein or the BD field equations
given in e qs.( 3) and (4), one can realize that the factors in
the metric functions,
\begin{eqnarray}
e^{2(k-U)}, ~~~W^2e^{2U}, ~~~h^2e^{-2U}
\end{eqnarray}
should have length dimension 2. Firstly, for the solution to the
Einstein field equations,
\begin{eqnarray}
e^{2(k_{E}-U_{E})} = L^2, ~~~W_{E}^2e^{2U_{E}} = 0, ~~~h_{E}^2e^{-2U_{E}} = L^2 \sin^2\theta
\end{eqnarray}
which all take the right length dimension, $2$, as expected.
Secondly, for the solution to the BD field equations, however, we
have
\begin{eqnarray}
e^{2(k_{BD}-U_{BD})} &=& L^2\left(\Delta \sin^{2}\theta \right)^{2/(2\omega+3)}, ~~~W_{BD}^2e^{2U_{BD}} = 0,  \\
h_{BD}^2e^{-2U_{BD}} &=& L^2 \sin^2\theta \left(\Delta \sin^{2}\theta \right)^{-2/(2\omega+3)}. \nonumber
\end{eqnarray}
Thus, in order for this metric solution to the BD field equations
to have the right length dimension, it turns out that the factor
$(\Delta \sin^{2}\theta )^{\pm 2/(2\omega+3)}$ should be
dimensionless. Therefore, in order to guarantee this, we introduce
a normalization factor $r_{0}$ to render the factor $\Delta $
(particularly appearing in $(\Delta \sin^{2}\theta )^{\pm
2/(2\omega+3)}$) dimensionless, i.e., $\Delta = [L^2(1 - 8\pi
\Lambda L^{2}/3) -2ML]/r^2_{0}$. The proper value of this
normalization factor $r_{0}$ shall be characterized later on when
we do the numerics to quantify the
rotation velocity in our model. \\

Then, by transforming back to the standard Schwarzschild
coordinates using Eq. (7), we finally arrive at the
Schwarzschild-de Sitter-type solution in the Brans-Dicke theory
given by
\begin{eqnarray}
ds^2 &=& \left(\Delta \sin^{2}\theta \right)^{-2/(2\omega+3)}
\left[- \left(1 - \frac{2\tilde{M}}{r} - \frac{8\pi \tilde{\Lambda}}{3}r^2\right)c^2 dt^2 + r^2 \sin^2 \theta d\phi^2 \right] \nonumber \\
&+& \left(\Delta \sin^{2}\theta \right)^{2/(2\omega+3)}
\left[\left(1 - \frac{2\tilde{M}}{r} - \frac{8\pi \tilde{\Lambda}}{3}r^2\right)^{-1}dr^2 + r^2 d\theta^2 \right],
\nonumber \\
\Phi (r, \theta) &=& \frac{1}{G_{0}}\left(\Delta \sin^{2}\theta
\right)^{2/(2\omega+3)},
\end{eqnarray}
where we have restored both the present value of Newton's constant
$G_{0}$ and the speed of light $c$ in order to come from the
geometrized unit ($G_{0} = c = 1$) back to CGS units. That is, we
have replaced $M$ with $G_{0}M/c^2 \equiv \tilde{M}$ along with
$\Lambda$$G_{0}\Lambda/c^4 \equiv \tilde{\Lambda}$;hence, $\Delta
= [r^2(1 - 8\pi \tilde{\Lambda} r^{2}/3) - 2\tilde{M}r]/r^2_{0}$.
A remarkable feature of this Schwarzschild-de Sitter-type solution
is the fact that, unlike the Schwarzschild-de Sitter solution in
general relativity, the spacetime it describes is static (i.e.,
non-rotating), but {\it not} spherically-symmetric. Note also that
as $\omega \rightarrow \infty$, this Schwarzschild-de Sitter-type
solution goes over to the standard Schwarzschild-de Sitter
solution in Einstein gravity as it should since the $\omega
\rightarrow \infty$ limit of the BD theory is the Einstein
gravity. Note also that the Brans-Dicke-Schwarzschild (BDS)
spacetime solution to these vacuum BD field equations has been
found some time ago \cite{hongsu1} in rather a theoretical attempt
to construct non-trivial black hole spacetime solutions in the BD
theory. There the BDS spacetime solution has been obtained by
setting $a = e = 0$ in the Brans-Dicke-Kerr-Newman (BDKN) solution
in Eq. (11) of Ref.9, and it coincides with the Schwarzschild-type
solution that results when we set $\tilde{\Lambda}=0$ in the
solution
constructed above, as it should. \\

Now coming back to our main objective, we are interested in the
role played by the BD scalar field as dark matter, particularly in
forming galactic dark matter halos inside of which the well-known
rotation curves have been observed. We suggest (and later on
demonstrate in terms of the reproduction of a flat rotation curve)
that, indeed, the BD scalar field can successfully play the role
of dark matter. In order for this to happen, we first claim that
the BD scalar can cluster into a halo-like configuration as it can
be represented by this Schwarzschild-de Sitter-type solution
constructed above in Eq. (14). At this point, therefore, it seems
relevant to address the nature of potential singularities of this
Schwarzschild-de Sitter-type metric solution. Just as the
Schwarzschild-de Sitter solution in general relativity, it appears
to possess two coordinate singularities which would arise at
$\Delta = [r^2(1 - 8\pi \tilde{\Lambda} r^{2}/3) -
2\tilde{M}r]/r^2_{0} = 0$ ; (i) the inner Schwarzschild
gravitational radius, and (ii) the outer de Sitter radius.
Generally, it is well-known that the Schwarzschild metric solution
in static coordinates possesses an event horizon, that the
solution essentially describes the region {\it outside} this event
horizon, and that its inside is not everywhere well-defined.
Meanwhile, the de Sitter metric solution in static coordinates has
a cosmological horizon,the solution mainly describes the region
{\it inside} this cosmological horizon, and its outside is not
everywhere well-defined. As such, the Schwarzschild-de Sitter
metric solution in static coordinates is able to represent the
regionbetween the inner event horizon and the outer cosmological
horizon, but {\it not} elsewhere. In the same spirit, our
Schwarzschild-de Sitter-type metric solution in the BD theory
given in static coordinates has been adopted here {\it only} to
represent the galactic dark halo region, but not below (i.e., the
interior of a given galaxy) nor beyond (i.e., the scale of galaxy
clusters or even of entire universe). Therefore, all we have to do
is to demonstrate that if we quantify the locations of these two
singularities by putting real numbers, they are well below and
beyond the typical domain of galactic halos.

First, we start with (i) the inner Schwarzschild gravitational
radius. For small-$r$, $\Delta = 0$ is approximated by $r -
2\tilde{M} \simeq 0$ whose solution is $r\simeq 2G_{0}M/c^2 \simeq
0.01 ~pc$. Here we used, for the typical (total) mass of a galaxy,
$M\sim 10^{11}M_{\odot}$. Obviously, this occurs well inside a
galaxy as the typical size of a galaxy ranges from few $kpc$ (for
dwarf galaxies) to few hundred $kpc$ (for ordinary galaxies).
Indeed, the physical meaning of this singularity is that if the
entire galaxy is squeezed into this gravitational radius, it
becomes a black hole with its event horizon placed at $0.01 ~pc$.
Namely, for ordinary galaxies, this inner Schwarzschild
gravitational radius is a {\it failed} Schwarzschild event
horizon. Next, we turn to (ii) the outer de Sitter radius. For
large-$r$, $\Delta = 0$ is approximated by $r - 8\pi
\tilde{\Lambda} r^{3}/3 \simeq 0$, whose solution is $r\simeq
\left(8\pi G_{0}\Lambda /3c^4 \right)^{-1/2} \simeq 4 ~Gpc$. Here
we used for the cosmological constant the observed value, $\Lambda
\simeq 10^{-8} ~erg/cm^3$. Apparently, this occurs well beyond a
single galaxy; Indeed, this is the scale of entire universe (given
the age of the universe, that is $\tau \sim 13.7 ~Gyr$, its rough
size would be $c\tau \sim 4 ~Gpc$). In other words, this outer de
Sitter radius is totally irrelevant.

To summarize, the potential singularities of the Schwarzschild-de
Sitter-type solution are irrelevant to keeping us from employing
the solution to represent the galactic dark halo region and,
hence, are harmless. Next, the seeming angular singularity at
$\theta = 0, ~\pi$ are irrelevant as well because the domain of
principal interest is the neighborhood of galactic equatorial
plane, $\theta = \pi/2$, inside the halo where most gases and
stars orbit around with flat rotation curves. Normally, the
symmetry axis is the last thing to be expected to be singular.
Indeed, close inspection reveals that the metric function factor
$(\Delta \sin^2 \theta)^{\pm 2/(2\omega +3)}$ is generated by the
solution-generating algorithm of Tiwari and Nayak \cite{tina} or
of Singh and Rai \cite{sinrai} which takes the (singularity-free)
solutions of the Einstein field equations to those of the BD field
equations. That is, since the appearance of the factor $(\sin^2
\theta)^{\pm 2/(2\omega +3)}$, which is responsible for the
singular nature of the symmetry axis, results {\it simply} from
the solution-generating algorithm, one naturally might expect that
the symmetry axis $\theta = 0, \pi$ cannot possibly be a genuine
curvature singularity as it would not really represent, say, an
infinite concentration of matter along there. However, the
symmetry axis $\theta = 0, ~\pi$, indeed, appears to be singular
as the invariant curvature polynomial, such as the curvature
scalar $R$, which can be readily calculated by contracting the
metric field equation and using the field equation for the BD
scalar field in Eq. (2), is given by
\begin{eqnarray}
R = \frac{4\omega}{(2\omega +3)^2}\frac{1}{r^2 \Delta}(\Delta \sin^2 \theta)^{-2/(2\omega +3)}
\left[r^2_{0} \Delta'^{2} + 4\left\{(2\omega +3)4\pi \Lambda G_{0}r^2 + \cot^2 \theta \right\}\Delta \right],  \nonumber
\end{eqnarray}
where $\Delta' = 2r\left[1-(\tilde{M}/r)-(16\pi
\tilde{\Lambda}r^2/3)\right]/r^2_{0}$, blows up there for generic
$\omega$ values. Namely, the symmetry axis appears to be a real
singularity rather than a coordinate singularity (that can be
removed by a change of coordinates).  This certainly is a very
unexpected and hence  puzzling feature of the Schwarzschild-de
Sitter-type solution in BD gravity theory. Therefore, one might
wonder what would happen to matter habiting near the galactic
poles when it comes close to the pole. We now hope to clarify the
true nature of this peculiar singularity along the symmetry axis
in some detail.

Among others, the first thing that comes to our mind regarding the
nature of this singularity along the symmetry axis is the fact
that, unlike the familiar curvature singularity at the center of a
black hole, which is point-like, this singularity is an
infinitely-extended line singularity. This means that perhaps the
fate of matter coming close to this singularity would be somewhat
different from what one would normally expect for matter
approaching a point-like black hole singularity. To get right to
the point, it turns out that in the {\it immediate vicinity} of
the symmetry axis, the specific energy (i.e., the energy per unit
mass) of a test particle becomes extremely high. As a result, the
particle moves along the symmetry axis at nearly the
ultrarelativistic speed (i.e., at nearly the speed of light). Of
course, this is an unusual feature, which has no analogue in the
Einstein gravity context, and a close inspection reveals that it
can eventually be attributed to the metric function factor
$(\Delta \sin^2 \theta)^{\pm 2/(2\omega +3)}$, which is
responsible for the singular nature of the symmetry axis. The
explicit demonstration of this peculiar behavior of test particles
near the symmetry axis shall be presented later on in section V in
terms of a rigorous analysis of the geodesic motion there.

In conclusion, this study of the true nature of the singularity
along the symmetry axis leads us to suspect that perhaps the
bizzare singularity at $\theta = 0, ~\pi$ of the Schwarzschild-de
Sitter-type solution in BD gravity theory can account for the
relativistic {\it bipolar outflows (twin jets)} extending from the
central region of ``active galactic nuclei (AGNs)'' Namely, the
curious singularity along the symmetry axis seems harmless, after
all. Instead, it turns out to be a pleasant surprise as it can
explain a long-known puzzle in observed features of galaxies. A
cautious comment might be needed here, though. That is, the
relativistic bipolar outflows have been observed only for some
particular types of galaxies, such as AGNs and micro-quasars.
(Radio galaxies, Seyfert galaxies and quasars fall into the AGN
category.) Namely, the jets do not seem to be a general
feature of all types of galaxies.   \\

Lastly, the failure of asymptotic flatness of this
Schwarzschild-de Sitter-type solution is not of serious concern
here as we essentially aim at finding a spacetime solution that
can represent the dark halo-like configuration, which is known to
cluster only on a sub-$Mpc$ scale, i.e., a local scale, and
eventually matches  the cosmological geometry, such as the
Friedmann-Robertson-Walker metric for homogeneous and isotropic
expanding universe, at a larger cosmological scale in order for
the BD theory in the presence of the cosmological constant to
provide a successful model for dark matter and dark energy, as suggested, for example, in Ref.4. \\
Our natural next mission is then to ask whether these
configurations really can reproduce the properties of dark matter
halos, namely, if our BD scalar model for dark matter can
reproduce the flattening of the rotation velocity curves inside
these halo configurations consistent with the observations.
Therefore, in the following, we shall address this issue;  first,
for the case when only the dark matter content represented by the
BD scalar is present, i.e., when the dark matter halo is
represented by the Schwarzschild-type solution (without the
cosmological constant ($\Lambda $) term), andthen for the other
case when both the dark matter content and the dark energy
content, represented by the cosmological constant, are present,
i.e., when the dark halo is represented by the Schwarzschild-de
Sitter-type solution (with the $\Lambda $ term).

\section{BD dark matter halos neglecting the dark energy ($\Lambda $) contribution: the case of the Jordan frame}

As claimed earlier, it appears that the BD scalar can indeed
cluster into a halo-like configuration as it can be represented by
the Schwarzschild-type solution in the BD theory. Thus, we now
attempt to obtain the rotation curves in our BD scalar halo and
eventually to demonstrate that they are actually flattened far out
to the distant region of halo. First of all, since we need
concrete ``numbers'' we now restore both Newton's constant $G_{0}$
and the speed of light $c$ in order to come from the geometrized
unit ($G_{0} = c = 1$) back to the CGS unit. Then, the
energy-momentum tensor of the BD scalar field given earlier in Eq.
(2) should now be multiplied by the factor
$\left(c^4/G_{0}\right)$, and in the Schwarzschild-type solution
in eEq. (14) above, we should replace $M$ by $G_{0}M/c^2 \equiv
\tilde{M}$, as we mentioned earlier. Apparently then, the
energy-momentum tensor of the BD scalar field with restored
$G_{0}$ and $c$ has the dimension of the energy-momentum density
in CGS units $(erg/cm^3)$.

We now turn to the computation of energy density profile and
(anisotropic) pressure components of the BD scalar field playing
the role of the dark matter by treating the BD scalar field as a
(dark matter) {\it fluid}. The BD scalar field fluid, however,
would fail to be a ``perfect'' fluid as can readily be envisaged
from the fact that the associated BDS solution configuration is
not spherically-symmetric. Namely, its pressure cannot be
``isotropic'', i.e., $P_{r} \neq P_{\theta} \neq P_{\phi}$. Such a
fluid may be called an {\it imperfect} fluid due to the {\it
anisotropic} pressure components, and as such, its stress tensor
can be written as
\begin{eqnarray}
T^{BD ~\mu}_{\nu} =
\pmatrix{-c{^2}\rho & 0 & 0 & 0 \cr
         0 & P_{r} & T^{r}_{\theta} & 0 \cr
         0 & T^{\theta}_{r} & P_{\theta} & 0 \cr
         0 & 0 & 0 & P_{\phi} \cr},
\end{eqnarray}
which is to be contrasted to its counterpart for the usual perfect
fluid with isotropic pressure given by the well-known form
$T^{\mu}_{\nu} = P\delta^{\mu}_{\nu} + (c^2 \rho +
P)U^{\mu}U_{\nu} = diag (-c^2 \rho, ~P, ~P, ~P)$, where
$U^{\alpha} = dX^{\alpha}/d\tau $ (with $\tau $ being the proper
time) denotes the 4-velocity of the fluid element normalized such
that $U^{\alpha}U_{\alpha} = -c^2$. Note that in addition to the
diagonal entries representing the (anisotropic) pressure
components $T^{i}_{i}=P_{i}$. (with no sum over $i$), there are
off-diagonal entries $T^{r}_{\theta}$, $T^{\theta}_{r}$
representing a {\it shear stress} that also results from the
failure of spherical symmetry.  It is also interesting to note
that a stress tensor of this sort (given in eEq. (15)) arises in
the case of rotating boson star made up of a complex scalar field
\cite{stress}. Thus by substituting the BDS solution given in Eq.
(14) into the BD energy-momentumm tensor in eq.(2) and then
setting it equal to Eq.(15), we can eventually read off the energy
density and the pressure components of the BD scalar field
imperfect fluid;
\begin{eqnarray}
\rho &=& {c^2\over 8\pi G_{0}}{4 \over (2\omega + 3)^2}{1\over r^2 \tilde{\Delta}}
\left(\Delta \sin^2 \theta\right)^{-2/(2\omega+3)}  \nonumber \\
&\times & \left[2(\omega +1)\left\{(r-\tilde{M})^2 + \tilde{\Delta} \cot^2 \theta \right\}
- (2\omega +3)\tilde{M}(r-\tilde{M})\right], \nonumber \\
P_{r} &=& -{c^4\over 8\pi G_{0}}{4 \over (2\omega + 3)^2}{1\over r^2 \tilde{\Delta}}
\left(\Delta \sin^2 \theta\right)^{-2/(2\omega+3)}   \nonumber \\
&\times& \left[(r-\tilde{M})^2 + (2\omega +3)\tilde{M}(2\tilde{M}-r) +
2(\omega -1)\tilde{\Delta} \cot^2 \theta \right], \nonumber \\
P_{\theta} &=& {c^4\over 8\pi G_{0}}{4 \over (2\omega + 3)^2}{1\over r^2 \tilde{\Delta}}
\left(\Delta \sin^2 \theta\right)^{-2/(2\omega+3)}
\left[(2\omega +3)(r-\tilde{M})(r-2\tilde{M})\right. \nonumber \\
&-& \left. 2(\omega -1)(r-\tilde{M})^2
+  \left\{2(\omega +1)\cos^2 \theta - (2\omega +3)\right\}{\tilde{\Delta} \over \sin^2 \theta }
\right], \\
P_{\phi} &=& -{c^4\over 8\pi G_{0}}{4 \over (2\omega + 3)^2}{1\over r^2 \tilde{\Delta}}
\left(\Delta \sin^2 \theta\right)^{-2/(2\omega+3)}  \nonumber \\
&\times& \left[2(\omega +1)(r-\tilde{M})^2 - \tilde{\Delta} \cot^2 \theta
- (2\omega +3)(r-\tilde{M})(r-2\tilde{M})\right], \nonumber \\
T^{r}_{\theta} &=& \Delta T^{\theta}_{r} =
{c^4\over 8\pi G_{0}}{4 \over (2\omega + 3)^2}{1\over r^2 }\cot \theta
\left(\Delta \sin^2 \theta\right)^{-2/(2\omega+3)}  \nonumber \\
&\times& \left[4\omega (r-\tilde{M}) - (2\omega
+3)(r-2\tilde{M})\right], \nonumber
\end{eqnarray}
where $\Delta = [r(r - 2\tilde{M})]/r^2_{0}$ and $\tilde{\Delta} =
r^2_{0}\Delta $. Note that the off-diagonal components
$T^{r}_{\theta}$ and $T^{\theta}_{r}$ are {\it odd} functions of
$\theta $ while the diagonal components $(\rho, P_{r}, P_{\theta},
P_{\phi})$ are {\it even} functions of the polar angle under
$\theta \rightarrow (\pi - \theta )$. As a result, the
off-diagonal components vanish (i.e., no shear stress survives) if
we average over this polar angle to get a net stress. Thus, the
equation of state of this BD scalar k-essence fluid forming a
galactic halo is given by
\begin{eqnarray}
w = {P\over c^2\rho } = -\frac{\left[(r-\tilde{M})^2 + (2\omega
+3)\tilde{M}(2\tilde{M}-r) + 2(\omega -1)\tilde{\Delta} \cot^2
\theta \right]} {2(\omega +1)\left\{(r-\tilde{M})^2 +
\tilde{\Delta} \cot^2 \theta \right\} - (2\omega
+3)\tilde{M}(r-\tilde{M})},
\end{eqnarray}
where $P = P_{r}$. Namely, $P = w(r, \theta)c^2 \rho $ with $w(r,
\theta) \sim O(1)$, meaning that this BD scalar fluid is
essentially a {\it barotropic} fluid but with a
``position-dependent'' coefficient $w(r, \theta)$. Note that
although the BD scalar field is a candidate for dark matter, it is
not quite a dust. In principle, the speed of sound in this BD
scalar field fluid can also be evaluated via $c^2_{s} = dP/d\rho$,
but we shall not discuss that in any more detail in this work.

We are now ready to compute the behavior of the rotation curves in
the outer region (i.e., at large, but finite$r$, say, $r >>
G_{0}M/c^2$ ) of our BD scalar field halo. To be more precise, for
a galaxy of typical (total) mass $M\sim 10^{11}M_{\odot}$, the
outer region of its dark matter halo, say, $r\sim 10 ~(kpc) \simeq
10^{23} ~(cm)$ is much greater than $G_{0}M/c^2 \simeq 10^{16}
~(cm)$ by a factor of $10^{7}$ or so. Thus, to this end, we first
approximate the expressions for its energy density and the
(radial) pressure given in Eq. (16) for large-$r$;
\begin{eqnarray}
\rho &\simeq & {c^2\over 8\pi G_{0}}{8(\omega + 1) \over (2\omega + 3)^2}{1\over r^2 \sin^2 \theta }
\Delta^{-2/(2\omega+3)}\sin^{-4/(2\omega+3)}\theta ,   \\
P &\simeq & -{c^4\over 2\pi G_{0}}{1 \over (2\omega + 3)^2}{1\over r^2}\left[2(\omega -1)\cot^2 \theta + 1\right]
\Delta^{-2/(2\omega+3)}\sin^{-4/(2\omega+3)}\theta . \nonumber
\end{eqnarray}
Note that in the above approximations and in the discussions
below, it was and it shall be assumed that the metric function
$\Delta = [r(r - 2\tilde{M})]/r^2_{0} \simeq (r/r_{0})^2$ for
large$r$. It is interesting to note that as a candidate for  dark
matter, the energy density $\rho $ of the BD scalar field is
almost certainly {\it positive everywhere} (i.e., for both small
and large$r$). In the meantime, its (radial) pressure $P$
particularly on a larger scale (i.e., for large$r$) turns out to
be {\it negative} although its sign appears unclear on a small
scale (i.e., for small$r$). \\

Finally, we are ready to determine the rotation curve inside our
BD scalar field halo. First, we start by recalling the origin of
the long-standing puzzle associated with the galaxy rotation
curves. In the most naive sense, the apparent rotation velocity of
an object at a radius $r$ from the galactic center would be given
in the Newtonian limit by  Kepler's third law, $v^2 =
G_{0}M(r)/r$. Indeed, inside a given galaxy where the luminous
mass is nearly uniformly distributed, the rotation velocity has
been observed to grow roughly linearly with the distance $r$,
consistent with this Kepler's law. In the outer region of a
galactic halo, however, the rotation velocity is expected to
behave like $v \sim 1/\sqrt{r}$ as the luminous mass is confined
within the extent of the given galaxy. The observations, however,
exhibit {\it flattened} rotation curves, and this has been the
age-old dilemma that called for the existence of dark matter
in the outer regions of galactic halos.   \\

In the present work, therefore, we shall employ the expression for
the rotation velocity still given by  Kepler's law, but instead
attempt to explain the flattened rotation curve in terms of the
non-trivial energy density of the BD scalar field, which appears
to play the role of dark matter. Namely, the non-trivial mass
density of the BD scalar field turns out to contribute to
the mass function $M(r)$ in such a way as to render the rotaion curve flat, as we shall see in a moment. \\

At this point, regarding the construction of the rotation
velocity, we have some comments. For the present case, it may seem
that one cannot just employ  Kepler's law as  the stress tensor of
the BD scalar field (playing the role of dark matter) and that the
background Schwarzschild-de Sitter-type spacetime in the BD theory
is not spherically symmetric. Nevertheless, gases or stars
orbiting around the host galaxy center usually lie on the galactic
equatorial plane. This implies that the orbital motions on the
equatorial plane $\theta = \pi/2$ are of particular interest;
hence, there one may still wish to employ Kepler's law. Along this
line, therefore, one might wish to employ the relativistic
counterpart to this rotation velocity equation, instead. Later on
in the appendix, the construction of the rotation velocity shall
be promoted to a fully relativistic version in terms of the
rigorous derivation of timelike geodesics in the Schwarzschild-de
Sitter-type spacetime. The quantitative results that we are about
to
present below, however, essentially remain unchanged even for the fully relativistic treatment.  \\

Now for our case, using the BD scalar energy density profile given
earlier, we have $M(r) = \int^{2\pi}_{0}d\phi \int^{\pi
-\epsilon}_{\epsilon}d\theta \int^{r}_{0}dr \sqrt{g_{rr}g_{\theta
\theta}g_{\phi \phi}}\rho (r, \theta) =
\left(2c^2/G_{0}\right)\left[(\omega +1)/(2\omega +1)(2\omega
+3)\right]f(\omega)r (r/r_{0})^{-2/(2\omega +3)}$ and hence,
\begin{eqnarray}
v^2(r) = \frac{G_{0}M(r)}{r} = c^2 \frac{2(\omega +1)}{(2\omega
+1)(2\omega +3)}f(\omega)
\left(\frac{r}{r_{0}}\right)^{-2/(2\omega +3)},
\end{eqnarray}
where $f(\omega) \equiv \int^{\pi -\epsilon}_{\epsilon}d\theta
\sin^{-[1+2/(2\omega +3)]}\theta = 2\int^{1-\delta}_{0}dx
[1-x^2]^{-(2\omega +4)/(2\omega +3)}$ with $\epsilon, ~\delta
<<1$. (Note here that the integration over the polar angle $\theta
$ starts not from $0$ but from $\epsilon <<1$ as the symmetry axis
$\theta = 0$ of the BDS solution in Eq. (14) possesses the danger
of an internal infinity nature, namely, the symmetry axis is an
infinite proper distance away, as discussed carefully in Ref.10.)
It has been known for some time that in order for the BD theory to
remain a viable theory of classical gravity passing all the
observational/experimental tests to date, the BD
$\omega$-parameter has to have a large value, say, $|\omega |\geq
500$ \cite{will}. In our previous study \cite{hongsu1}, however,
we realized that the static solution to the vacuum BD field
equations given in Eq. (14) above, but without $\Lambda $, can
turn into a black hole spacetime for $-5/2 \leq \omega <-3/2$.
Thus, now for $|\omega |\geq 500$, the same static solution Eq.
(14) we are considering represents just a halo-like configuration
with a regular geometry everywhere (i.e., having no horizon),
which is static but not exactly spherically symmetric (note that
the galactic halos are also believed to be nearly spherically
symmetric, but not exactly). Thus, if we substitute a
large-$\omega$ value, say, $\omega \sim 10^6$ into Eq. (19) above,
evidently $M(r) \sim r$; hence, we get
\begin{eqnarray}
v(r) \simeq 100 (km/s) \times \left(\frac{r}{r_{0}}\right)^{-(1/10^6)}
\end{eqnarray}
because for $\omega \sim 10^6$, $f(\omega) \simeq O(1)$. Namely,
for this large-$\omega$ value, the rotation curve gets flattened
out as $r^{-(10^{-6})} \sim constant $, and its magnitude becomes
several hundred $km/s$. Here, it is particularly remarkable that
the observationally and experimentally allowed large value of
$\omega \sim 10^6$ renders the rotation curve flat and fixes its
magnitude to several hundred $km/s$ at the {\it same} time! This
is, indeed, an attractive feature of the BD theory that
distinguishes it from other theories of dark matter proposed thus
far. Of course, this is in impressive agreement with the data for
rotation curves observed in spiral/elliptic galaxies with $M/L
\simeq (10 - 20) M_{\odot}/L_{\odot}$ and in
low-surface-brightness (LSB)/dwarf galaxies with $M/L \simeq (200
- 600) M_{\odot}/L_{\odot}$ (where $M/L$ denotes the so-called
``mass-to-light'' ratio given in the units of the solar
mass-to-luminosity ratio and exhibiting a large excess of dark
matter over the luminous matter) \cite{rc}. For instance, the
rotation curve of the dwarf spiral galaxy $M33$ is shown in Fig.1.
\begin{figure}[hbt]
\centerline{\epsfig{file=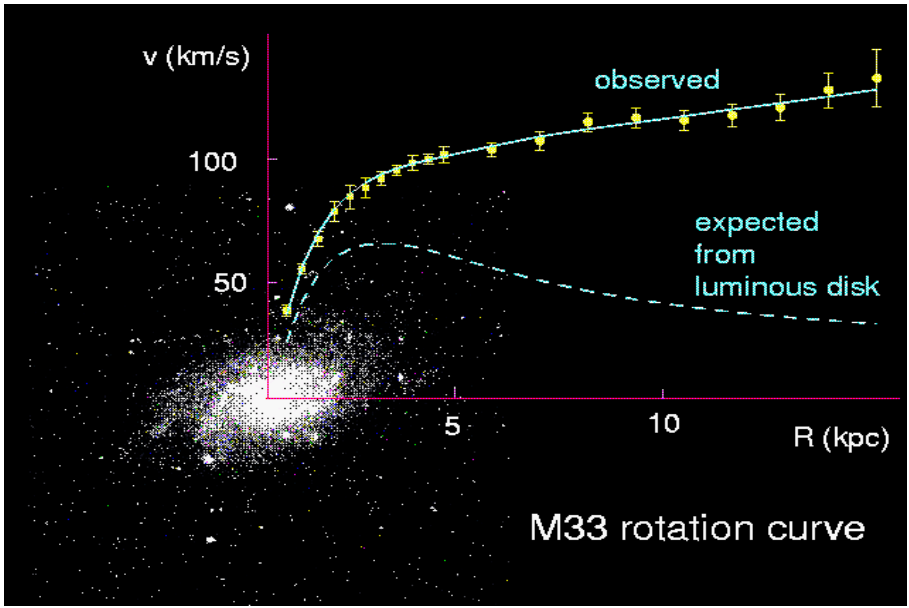, width=12cm, height=8cm}}
\caption{\ Flattened rotation curve in the dark halo of dwarf spiral galaxy $M33$.}
\end{figure}
Rotation curves are observed usually via  measurements of the
Doppler shift of the $21-cm$ emission line from neutral hydrogen
(HI) for distant galaxies and of the light emitted by stars of
nearby galaxies \cite{halo1, halo2}. It is also interesting to
note that this behavior of the rotation curve in our BD theory
dark matter halo model is {\it independent} of the mass of the
host galaxy. Namely, this behavior of the rotation curve comes
exclusively from the nature of the dark matter, i.e., the BD
scalar field. Indeed, we have restricted our interest in the
present work only to the contribution coming from the dark matter
component, i.e., the BD scalar field and {\it not} to the combined
contributions from both the luminous and the dark matter.
Therefore, the resulting galactic rotation curve (having a
contribution coming {\it only} from the BD scalar, which is the
dark matter candidate) comes out as being independent of the
luminous mass in each galaxy. The well-known Tully-Fisher relation
\cite{tf} between the {\it total} luminosity of a given galaxy and
the outermost rotation velocity can be accounted for if both the
luminous and the dark matter contributions are considered. This
issue has not been addressed in the present work, but hopefully it
should be answered by any dark matter model
in order for the proposed model to be truly successful. \\

We also point out that even if we employ more exact expression for
the rotation velocity curve involving the Doppler shift of light
emitted by the orbiting objects (assuming that the k-essence halo
is almost sphericallysymmetric), namel,y $v^2(r) = G_{0}M(r)/r +
4\pi r^2 G_{0}P/c^2$ \cite{khalo} (with $P$ being the radial
pressure given in Eq. (18) above), the conclusions above remain
the same. Thus, in what follows, we shall demonstrate this in some
detail. Using Eqs. (18) and (19), we have
\begin{eqnarray}
&&v^2(r) = \frac{G_{0}M(r)}{r} + \frac{4\pi G_{0}}{c^2} r^2 P  \\
&&= c^2 \frac{2(\omega +1)}{(2\omega +1)(2\omega +3)}f(\omega)\left(\frac{r}{r_{0}}\right)^{-2/(2\omega +3)}
- c^2 \frac{2[2(\omega -1)\cot^2 \theta +1]}{(2\omega +3)^2}\sin^{-4/(2\omega+3)}\theta
\left(\frac{r}{r_{0}}\right)^{-4/(2\omega +3)}. \nonumber
\end{eqnarray}
Again, due to the failure of spherical symmetry in the BDS
solution, Eq.(14), which is identified with a dark matter halo
configuration, this rotation velocity comes out to be polar angle
($\theta $) dependent. However since the gases or the stars
orbiting around the galaxy center usually lie on the galactic
(equatorial) plane, the relevant situation to consider is the case
$\theta = \pi/2$ in which
\begin{eqnarray}
v(r) &=& c \left[\frac{2(\omega +1)}{(2\omega +1)(2\omega +3)}f(\omega)\left(\frac{r}{r_{0}}\right)^{-2/(2\omega +3)}
- \frac{2}{(2\omega +3)^2}\left(\frac{r}{r_{0}}\right)^{-4/(2\omega +3)}\right]^{1/2} \nonumber \\
&\simeq & 100 (km/s) \times
\left(\frac{r}{r_{0}}\right)^{-(1/10^6)},
\end{eqnarray}
where again we have chosen the value $\omega \sim 10^6$ in the last line.

Next, the equation of state in eEq. (17) of this BD scalar field
fluid becomes, in the outer region of the galactic dark matter
halo (i.e., at large$r$),
\begin{eqnarray}
w \simeq - \frac{\left[2(\omega -1)\cos^2 \theta + \sin^2 \theta \right]}
{2(\omega + 1)}
\end{eqnarray}
which is obviously negative due to the {\it negative} pressure
(and still {\it positive} energy density) in this outer region.
Moreover, for a large-$\omega$ value, i.e., $\omega \sim 10^6$,
for which the rotation curve gets flattened out, that we just have
realized, this equation of state at large$r$ further approaches $w
\simeq - \cos^2 \theta \simeq - O(1)$. (Incidentally, it is
interesting to note that in the vicinity of the equatorial plane
$\theta = \pi/2$, $w = 0$; namely, the BD scalar behaves  nearly
as dust.) This observation is particularly interesting as it
appears to indicate that the BD scalar field we are considering
possesses {\it dark -energy-like} negative pressure on larger
scales. This observation is indeed consistent with our previous
study \cite{hongsu2}, which showed that on a cosmological scale,
the BD scalar field, which has been identified with a
``k-essence'' there, did exhibit the nature of dark energy
possessing a negative pressure.

\section{BD dark matter halos including the dark energy ($\Lambda $) contribution}

By substituting the Schwarzschild-de Sitter-type solution given in
Eq. (14),, but this time into the {\it total} energy-momentum
tensor including the contribution coming from the cosmological
constant term
\begin{eqnarray}
T_{\mu \nu} &=& T^{BD}_{\mu \nu} - \frac{1}{G_{0}\Phi}\Lambda g_{\mu \nu} \\
&=& {c^4\over 8\pi G_{0}}\left[{\omega \over
\Phi^2}(\nabla_{\mu}\Phi \nabla_{\nu}\Phi - {1\over 2}g_{\mu
\nu}\nabla_{\alpha}\Phi \nabla^{\alpha}\Phi) + {1\over
\Phi}(\nabla_{\mu} \nabla_{\nu}\Phi - g_{\mu
\nu}\nabla_{\alpha}\nabla^{\alpha}\Phi)\right]  -
\frac{1}{G_{0}\Phi}\Lambda g_{\mu \nu}, \nonumber
\end{eqnarray}
and then again setting it equal to Eq.(15), we can eventually read
off the energy density and the pressure components of the BD
scalar field imperfect fluid as
\begin{eqnarray}
\rho &=& {c^2\over 8\pi G_{0}}{4 \over (2\omega + 3)^2}{1\over r^2 \tilde{\Delta}}
\left(\Delta \sin^2 \theta\right)^{-2/(2\omega+3)} \left[\left\{(r-\tilde{M})-\frac{16\pi}{3}\tilde{\Lambda}r^3\right\} \right. \nonumber \\
&\times & \left. \left[2\omega \left\{(r-\tilde{M})-\frac{16\pi}{3}\tilde{\Lambda}r^3\right\} +
\left\{2(r-\tilde{M})-(2\omega +3)\tilde{M}+(2\omega -1)\frac{8\pi}{3}\tilde{\Lambda}r^3\right\}\right] \right. \nonumber \\
&+& \left. 2(\omega +1)\tilde{\Delta} \cot^2 \theta \right] + \frac{(2\omega -1)}{(2\omega +3)}\frac{\Lambda }{c^2}\left(\Delta \sin^2 \theta\right)^{-2/(2\omega+3)},
\nonumber \\
P_{r} &=& -{c^4\over 8\pi G_{0}}{4 \over (2\omega + 3)^2}{1\over r^2 \tilde{\Delta}}
\left(\Delta \sin^2 \theta\right)^{-2/(2\omega+3)} \left[\left\{(r-\tilde{M})-\frac{16\pi}{3}\tilde{\Lambda}r^3\right\} \right. \nonumber \\
&\times& \left. \left[2(\omega +1)\left\{(r-\tilde{M})-\frac{16\pi}{3}\tilde{\Lambda}r^3\right\} +
\left\{2(r-\tilde{M})-(2\omega +3)\tilde{M}+(2\omega -1)\frac{8\pi}{3}\tilde{\Lambda}r^3\right\}\right] \right. \nonumber \\
&-& \left. (2\omega +3)\tilde{\Delta} (1-16 \pi \tilde{\Lambda}r^2) + 2(\omega -1)\tilde{\Delta} \cot^2 \theta \right]
- \frac{(2\omega -1)}{(2\omega +3)}\Lambda \left(\Delta \sin^2 \theta\right)^{-2/(2\omega+3)},
\nonumber \\
P_{\theta} &=& {c^4\over 8\pi G_{0}}{4 \over (2\omega + 3)^2}{1\over r^2 \tilde{\Delta}}
\left(\Delta \sin^2 \theta\right)^{-2/(2\omega+3)} \nonumber \\
&\times& \left[\left\{(r-\tilde{M})-\frac{16\pi}{3}\tilde{\Lambda}r^3\right\}
\left[(2\omega +3)\frac{\tilde{\Delta}}{r} - 2(\omega -1)\left\{(r-\tilde{M})-\frac{16\pi}{3}\tilde{\Lambda}r^3\right\}
\right] \right. \nonumber \\
&+& \left. \left\{2(\omega +1)\cos^2 \theta - (2\omega +3)\right\}\frac{\tilde{\Delta} }{\sin^2 \theta} \right]
- \frac{(2\omega -1)}{(2\omega +3)}\Lambda \left(\Delta \sin^2 \theta\right)^{-2/(2\omega+3)}, \\
P_{\phi} &=& -{c^4\over 8\pi G_{0}}{4 \over (2\omega + 3)^2}{1\over r^2 \tilde{\Delta}}
\left(\Delta \sin^2 \theta\right)^{-2/(2\omega+3)}  \nonumber \\
&\times& \left[\left\{(r-\tilde{M})-\frac{16\pi}{3}\tilde{\Lambda}r^3\right\}
\left[2(\omega +1)\left\{(r-\tilde{M})-\frac{16\pi}{3}\tilde{\Lambda}r^3\right\} - (2\omega +3)\frac{\tilde{\Delta}}{r}
\right] \right. \nonumber \\
&-& \left. \tilde{\Delta} \cot^2 \theta \right]
- \frac{(2\omega -1)}{(2\omega +3)}\Lambda \left(\Delta \sin^2 \theta\right)^{-2/(2\omega+3)},
\nonumber \\
T^{r}_{\theta} &=& \Delta T^{\theta}_{r} =
{c^4\over 8\pi G_{0}}{4 \over (2\omega + 3)^2}{1\over r^2 }\cot \theta
\left(\Delta \sin^2 \theta\right)^{-2/(2\omega+3)}  \nonumber \\
&\times& \left[4\omega \left\{(r-\tilde{M}) - \frac{16\pi}{3}\tilde{\Lambda}r^3\right\} - (2\omega +3)\frac{\tilde{\Delta}}{r}\right]
\nonumber
\end{eqnarray}
where now $\Delta = [r^2(1 - 8\pi \tilde{\Lambda} r^{2}/3) - 2\tilde{M}r]/r^2_{0}$ and $\tilde{\Delta} = r^2_{0}\Delta $.
As in the previous case where the contribution from the dark energy component, namely, $\Lambda $ has been neglected,
the off-diagonal components $T^{r}_{\theta}$, $T^{\theta}_{r}$ are {\it odd} functions
of $\theta $ whereas the diagonal components $(\rho, P_{r}, P_{\theta}, P_{\phi})$ are {\it even}
functions of the polar angle under $\theta \rightarrow (\pi - \theta )$. Consequently, the off-diagonal
components vanish (i.e., no shear stress survives) if we average over this polar angle to get a net
stress.
Again, first the equation of state of this BD scalar k-essence fluid forming a galactic halo is given by
\begin{eqnarray}
w = {P\over c^2\rho }.
\end{eqnarray}
Equations (25) and (26) indicate that $P = w(r, \theta)c^2 \rho $
with $w(r, \theta) \sim O(1)$ meaning that this BD scalar fluid is
still a {\it barotropic} fluid but with ``position-dependent''
coefficient $w(r, \theta)$.

As before, we now turn to the exploration of the behavior of
rotation curves in the outer region (i.e., at large but
finite-$r$, say, $r\sim 10 ~(kpc) \simeq 10^{23} ~(cm) >>
G_{0}M/c^2 \simeq 10^{16} ~(cm)$) of our BD scalar field halo. And
to this end, we first approximate the expressions for the energy
density and the (radial) pressure of the BD scalar fluid given in
eq.(25) using essentially $\Delta = [r^2(1 - 8\pi \tilde{\Lambda}
r^{2}/3) - 2\tilde{M}r]/r^2_{0} \simeq [r^2(1 - 8\pi
\tilde{\Lambda} r^{2}/3)]/r^2_{0}$ for large-$r$. They are
\begin{eqnarray}
\rho &\simeq & {c^2\over 8\pi G_{0}}{8(\omega + 1) \over (2\omega + 3)^2}
{1\over r^2 \sin^2 \theta }\left(\Delta \sin^2 \theta\right)^{-2/(2\omega+3)} \nonumber \\
&+& \left[\frac{(2\omega -1)}{(2\omega +3)} - \frac{4}{3}\frac{(2\omega +1)}{(2\omega +3)^2}\right]\frac{\Lambda}{c^2}
\left(\Delta \sin^2 \theta\right)^{-2/(2\omega+3)},   \\
P &\simeq & -{c^4\over 2\pi G_{0}}{1 \over (2\omega + 3)^2}{1\over r^2}\left[2(\omega -1)\cot^2 \theta + 1\right]
\left(\Delta \sin^2 \theta\right)^{-2/(2\omega+3)} \nonumber \\
&-& \left[\frac{(2\omega -1)}{(2\omega +3)} + \frac{4}{3}\frac{1}{(2\omega +3)^2}\right]\Lambda
\left(\Delta \sin^2 \theta\right)^{-2/(2\omega+3)}. \nonumber
\end{eqnarray}
It is interesting to note that the energy
density $\rho $ of the BD scalar field even in the presence of $\tilde{\Lambda}$, the dark energy component,
is almost certainly {\it positive everywhere} (i.e., for both small and large-$r$).
In the mean time, its (radial) pressure $P$, in the presence of $\tilde{\Lambda}$, particularly at larger scale
(i.e., for large-$r$) turns out to be {\it negative}. \\
We are now ready to determine the rotation curve inside our BD scalar field halo.
First, using the BD scalar field energy density profile given earlier, we are supposed to compute the {\it mass function}
$M(r) = \int^{2\pi}_{0}d\phi
\int^{\pi -\epsilon}_{\epsilon}d\theta \int^{r}_{0}dr \sqrt{g_{rr}g_{\theta \theta}g_{\phi \phi}}\rho (r, \theta)$
in which we have $\sqrt{g_{rr}g_{\theta \theta}g_{\phi \phi}} = \left(\Delta \sin^2 \theta\right)^{1/(2\omega+3)}
\Delta^{-1/2}(r^3/r_{0})\sin \theta $. Although we shall consistently employ the approximation $\Delta \simeq [r^2(1 - 8\pi \tilde{\Lambda} r^{2}/3)]/r^2_{0}$
as we are interested in the behavior of rotation curves in the outer region (i.e., at large-$r$) of the BD scalar halo,
the actual computation of the mass function $M(r)$ is unfortunately not available since the analytic integration over $r$ cannot be done in a closed form.
Indeed one of our main objectives in this section, where the contribution from the dark energy component (i.e., $\Lambda $) has been included, involves
the theoretical derivation of, say, the upper bound on the value of the cosmological constant $\Lambda $ that would result in, despite its presence,
the nearly flattened rotation curves as we observe them. Obviously, the (positive) cosmological constant has positive contribution to the mass-energy
density $\rho $ and as a result, in the presence of $\Lambda $, the mass function $M(r)$ grows {\it non-linearly} with the distance (i.e., the radius of the dark halo)
ruining the asymptotic flatness of the rotation curves. As such, in order for our model, i.e., the BD theory with (positive) cosmological constant to
successfully reproduce the flattened rotation curves, it is evident that the absolute magnitude of the cosmological constant should be small enough.
For this reason, we shall, for the sake of explicit, analytic computation of the mass function $M(r)$, assume that the dimensionless quantity,
$\tilde{\Lambda} r^{2}$ be very small, i.e., $(G_{0}\Lambda/c^4) r^2 << 1$ and see what we would end up with. Then we have
\begin{eqnarray}
M(r) &\simeq& \int^{2\pi}_{0}d\phi \int^{\pi -\epsilon}_{\epsilon}d\theta \int^{r}_{0}dr
\left\{{c^2\over 8\pi G_{0}}{8(\omega + 1) \over (2\omega + 3)^2}
{r\over \sin \theta }
+ \left[\frac{(2\omega -1)}{(2\omega +3)} - \frac{4}{3}\frac{(2\omega +1)}{(2\omega +3)^2}\right]\frac{\Lambda}{c^2}
r^3 \sin \theta \right\} \nonumber \\
&\times& \Delta ^{-1/2}\left(\Delta \sin^2 \theta\right)^{-1/(2\omega+3)} \nonumber \\
&\simeq& \frac{c^2}{G_{0}}\frac{2(\omega +1)}{(2\omega +3)^2}f(\omega ) \\
&\times& \left[\frac{(2\omega +3)}{(2\omega +1)}r + \frac{(2\omega +5)}{(6\omega +7)}\frac{4\pi G_{0}}{3c^4}\Lambda r^3 +
\frac{2}{(10\omega +13)}\left(\frac{4\pi G_{0}}{3c^4}\Lambda \right)^2 r^5\right]\left(\frac{r}{r_{0}}\right)^{-2/(2\omega +3)} \nonumber \\
&+& 2\pi \left\{\frac{(2\omega -1)}{(2\omega +3)} - \frac{4(2\omega +1)}{3(2\omega +3)^2}\right\}\frac{\Lambda}{c^2}g(\omega) \nonumber \\
&\times& \left[\frac{(2\omega +3)}{(6\omega +7)}r^3 + \frac{(2\omega +5)}{(10\omega +13)}\frac{4\pi G_{0}}{3c^4}\Lambda r^5 +
\frac{2}{(14\omega +19)}\left(\frac{4\pi G_{0}}{3c^4}\Lambda \right)^2 r^7\right]\left(\frac{r}{r_{0}}\right)^{-2/(2\omega +3)} \nonumber
\end{eqnarray}
where now $f(\omega) \equiv \int^{\pi -\epsilon}_{\epsilon}d\theta \sin^{-[1+2/(2\omega +3)]}\theta
= 2\int^{1-\delta}_{0}dx [1-x^2]^{-(2\omega +4)/(2\omega +3)} \simeq O(1)$ and
$g(\omega) \equiv \int^{\pi -\epsilon}_{\epsilon}d\theta \sin^{[1-2/(2\omega +3)]}\theta
= 2\int^{1-\delta}_{0}dx [1-x^2]^{-1/(2\omega +3)} \simeq O(1)$ with $\epsilon, ~\delta <<1$.
And here we used $\Delta ^{-1/2}\Delta ^{-1/(2\omega+3)} \simeq \left(1 + 4\pi G_{0}\Lambda r^2/3c^4\right)
\left(1 + 8\pi G_{0}\Lambda r^2/3(2\omega +3)c^4\right)r^{-1}(r/r_{0})^{-2/(2\omega +3)}$ since we assumed $(G_{0}\Lambda/c^4) r^2 << 1$ as explained above.
Lastly, the rotation velocity is given by
\begin{eqnarray}
v^2(r) &=& \frac{G_{0}M(r)}{r} = c^2 \frac{2(\omega +1)}{(2\omega +1)(2\omega +3)}f(\omega)\left(\frac{r}{r_{0}}\right)^{-2/(2\omega +3)} \\
&+& \left[\left\{\left[\frac{(2\omega -1)}{(6\omega +7)} - \frac{4(2\omega +1)}{3(2\omega +3)(6\omega +7)}\right]\frac{2\pi G_{0}}{c^2}\Lambda g(\omega) +
\frac{2(\omega +1)(2\omega +5)}{(2\omega +3)^2(6\omega +7)}\frac{4\pi G_{0}}{3c^2}\Lambda f(\omega)\right\} r^2  \right. \nonumber \\
&+& \left. c^2\left\{\frac{4(\omega +1)}{(2\omega +3)^2(10\omega +13)}\left(\frac{4\pi G_{0}}{3c^4}\Lambda \right)^2 f(\omega) \right. \right. \nonumber \\
&+& \left. \left. \left[\frac{(2\omega -1)}{(2\omega +3)} - \frac{4(2\omega +1)}{3(2\omega +3)^2}\right]\frac{(2\omega +5)}{(10\omega + 13)}
\frac{3}{2}\left(\frac{4\pi G_{0}}{3c^4}\Lambda \right)^2 g(\omega)\right\}r^4 \right. \nonumber \\
&+& \left. c^2\left\{\left[\frac{(2\omega -1)}{(2\omega +3)} - \frac{4(2\omega +1)}{3(2\omega +3)^2}\right]\frac{3}{(14\omega + 19)}\left(\frac{4\pi G_{0}}{3c^4}\Lambda \right)^3 g(\omega)\right\}r^6\right]
\left(\frac{r}{r_{0}}\right)^{-2/(2\omega +3)}. \nonumber
\end{eqnarray}
Thus again if we substitute a large-$\omega$ value, say, $\omega \sim 10^6$ into eq.(29) above, we get, as $f(\omega), g(\omega)\sim O(1)$,
\begin{eqnarray}
v^2(r) &\simeq& \left[100 (km/s) \times \left(\frac{r}{r_{0}}\right)^{-(1/10^6)}\right]^2 \\
&+& \left[\frac{2\pi}{3}\left(\frac{G_{0}}{c^2}\Lambda r^2_{0}\right)\left(\frac{r}{r_{0}}\right)^2 +
c^2\frac{3}{10}\left(\frac{4\pi G_{0}}{3c^4}\Lambda r^2_{0}\right)^2 \left(\frac{r}{r_{0}}\right)^4\right]\left(\frac{r}{r_{0}}\right)^{-(1/10^6)}. \nonumber
\end{eqnarray}
where we introduced the normalization factor $r_{0} \sim 10 ~(kpc) \simeq 10^{23} ~(cm)$ which represent a typical distance to the outer region of the dark halo where
rotation curves begin to get flattened. We are now in a position to {\it theoretically} determine the upper bound on the value of the (positive) cosmological constant
in our dark matter model, namely the BD theory in the presence of the cosmological constant. That is, in order for this result in eq.(30) to successfully describe
the flattened rotation curve in the outer region of dark halo, $r \geq r_{0}$, it should be imposed that
\begin{eqnarray}
&(A)& ~~~\frac{G_{0}\Lambda}{c^2}r^2_{0} << (100 km/s)^2, \nonumber \\
&(B)& ~~~\left(\frac{G_{0}\Lambda}{c^4}r^2_{0}\right)^2 << 10^{-6} \nonumber
\end{eqnarray}
as $\left[G_{0}\Lambda /c^2\right] = 1/s^2$ and $\left[G_{0}\Lambda /c^4\right] = 1/cm^2$.
Then using $G_{0} = 6.67\times 10^{-8} ~(cm^3/g s^2)$ and $c = 3\times 10^{10} ~(cm/s)$, the conditions $(A)$ and $(B)$ give
$\Lambda << 1.35 \times 10^{-4} ~(erg/cm^3)$. It thus is interesting to remark that this {\it theoretical} upper bound on the value of the cosmological constant {\it is}
consistent with the observed value $\Lambda_{obs} \simeq 10^{-8} ~(erg/cm^3)$ (or $\Lambda_{obs}/c^2 \simeq 10^{-29} ~(g/cm^3)$) \cite{wmap, lambda}.
Of course an alternative interpretation is acceptable as well. That is, with the observed small value $\Lambda_{obs} \simeq 10^{-8} ~(erg/cm^3)$,
our model for dark matter turns out to be able to reproduce the flattened rotation curve in a successful manner.
Therefore, it appears that our model, based on the Brans-Dicke theory in the presence of the cosmological constant, can successfully reproduce both the
dark matter halo configuration and the flattened rotation curves inside of it. This may indicate that our model could be one of the promising candidates for
dark matter but other known evidences for dark matter need to be tested as well in this context of BD theory (possibly with cosmological constant)
in order for it to be truly successful model of dark matter (and dark energy). \\
Next, the equation of state in eq.(26) of this BD scalar field becomes,
in the outer region of the galactic dark matter halo (i.e., at large-$r$),
\begin{eqnarray}
w &\simeq& - \frac{\left[2(\omega -1)\cos^2 \theta + \sin^2 \theta \right]
+ \left[(2\omega - 1)(2\omega + 3) + 4/3\right]\left(2\pi G_{0}/c^4\right)\Lambda r^2 \sin^2 \theta}
{2(\omega + 1) + \left[(2\omega -1)(2\omega + 3) - 4(2\omega +1)/3\right]\left(2\pi G_{0}/c^4\right)\Lambda r^2 \sin^2 \theta}  \nonumber \\
&\simeq& - \frac{\left[(2\omega - 1)(2\omega + 3) + 4/3\right]}
{\left[(2\omega - 1)(2\omega + 3) - 4(2\omega +1)/3\right]}
\end{eqnarray}
which is obviously negative due to the {\it negative} pressure (and still {\it positive}
energy density) in this outer region. Moreover, for the large-$\omega$ value,
i.e., $\omega \sim 10^6$ for which the rotation curve gets flattened out, this equation of state at large-$r$ further approaches
$w \simeq - 1$. Once again, this observation is particularly interesting as
it appears to indicate that the BD scalar we are considering possesses
{\it dark energy-like} negative pressure on larger, cosmological scales
consistently with our previous study \cite{hongsu2} that on the cosmological scale, the BD
scalar field with $\Lambda $ does exhibit the nature of dark energy possessing the negative pressure.

\section{Rigorous relativistic treatment of the rotation velocity and the bipolar jets}

\subsection{The Rotation Velocity - a Fully Relativistic Derivation}

In the previous section, the rotation velocity of a test body orbiting in the background of the Schwarzschild-de Sitter-type spacetime in BD theory has been
taken as the expression in the Newtonian limit approximation. In the present section, the construction of the rotation velocity shall
be promoted to a fully relativistic version in terms of the rigorous derivation of timelike geodesics in the Schwarzschild-de Sitter-type spacetime. \\
The Schwarzschild-de Sitter-type spacetime in BD theory given in eq.(14) in the text is static and axisymmetric. Thus it possesses time-translational
isometry generated by the timelike Killing vector $\xi^{\mu} = \left(\partial /\partial t\right)^{\mu} = \delta^{\mu}_{t}$ and rotational isometry
generated by the axial Killing vector $\psi^{\mu} = \left(\partial /\partial \phi\right)^{\mu} = \delta^{\mu}_{\phi}$. And the associated conserved
quantities are energy $\tilde{E}=E/m $ and angular momentum $\tilde{L}=L/m $ (per unit rest mass) of a test body along the geodesic
\begin{eqnarray}
-\tilde{E} &=& u_{\mu}\xi^{\mu} = g_{\alpha t}u^{\alpha} =
- \left(\Delta \sin^2 \theta\right)^{-2/(2\omega+3)}\left(1 - \frac{2\tilde{M}}{r} - \frac{8\pi \tilde{\Lambda}}{3}r^2\right)c^2\left(\frac{dt}{d\tau}\right), \\
\tilde{L} &=& u_{\mu}\psi^{\mu} = g_{\alpha \phi}u^{\alpha} =
\left(\Delta \sin^2 \theta\right)^{-2/(2\omega+3)}r^2 \sin^2 \theta \left(\frac{d\phi }{d\tau}\right)
\end{eqnarray}
where $u^{\mu} = \left(dx^{\mu}/d\tau \right)$ denotes the 4-velocity which is tangent to the geodesic. In addition, we have $g_{\mu \nu}u^{\mu}u^{\nu} = -\kappa $,
namely
\begin{eqnarray}
-\kappa &=& \left(\Delta \sin^{2}\theta \right)^{-2/(2\omega+3)}
\left[- \left(1 - \frac{2\tilde{M}}{r} - \frac{8\pi \tilde{\Lambda}}{3}r^2\right)c^2 \left(\frac{dt}{d\tau}\right)^2 + r^2 \sin^2 \theta \left(\frac{d\phi}{d\tau}\right)^2 \right] \nonumber \\
&+& \left(\Delta \sin^{2}\theta \right)^{2/(2\omega+3)}
\left[\left(1 - \frac{2\tilde{M}}{r} - \frac{8\pi \tilde{\Lambda}}{3}r^2\right)^{-1}\left(\frac{dr}{d\tau}\right)^2 + r^2 \left(\frac{d\theta}{d\tau}\right)^2 \right].
\end{eqnarray}
where $\kappa = c^2$ for timelike geodesics (i.e., for massive bodies) and $\kappa = 0$ for null geodesics (i.e., for light rays). Now, one may use eqs.(32) and (33) to eliminate
$\left(dt/d\tau \right)$ and $\left(d\phi/d\tau \right)$ in terms of $\tilde{E}$ and $\tilde{L}$ and the result may be substituted into eq.(34) in order to obtain the
``first integral'' of the radial geodesic equation. Further, since gases or stars orbiting around the host galaxy center usually lie on the galactic (equatorial) plane, the case of
equatorial geodesics at $\theta = \pi/2$ is of particular interest and the result is
\begin{eqnarray}
\left(\frac{dr}{d\tau}\right)^2 &+& V_{eff}(\tilde{E}, \tilde{L} ; r) = 0, \\
V_{eff}(\tilde{E}, \tilde{L} ; r) &=& \left(1 - \frac{2\tilde{M}}{r} - \frac{8\pi \tilde{\Lambda}}{3}r^2\right)
\left[\frac{\tilde{L}^2}{r^2} + \kappa \Delta^{-2/(2\omega+3)}\right] - \frac{\tilde{E}^2}{c^2}. \nonumber
\end{eqnarray}
Therefore, now the problem of obtaining the timelike and null geodesics on the equatorial plane of the Schwarzschild-de Sitter-type spacetime in BD theory reduces to solving
a problem of non-relativistic, one-dimensional motion in an effective potential $V_{eff}(\tilde{E}, \tilde{L} ; r)$. Here we consider the case of our particular interest,
the ``stable'' circular orbit motions of massive objects characterized by the simultaneous conditions of
\begin{eqnarray}
V_{eff}  = 0, ~~~\frac{dV_{eff}}{dr} = 0
\end{eqnarray}
as the ones first studied by Bardeen, Press and Teukolsky \cite{bpt} for the case of Kerr spacetime. Then the conditions for the stable circular orbit motions in eq.(36) yield
the required values of the specific energy and the specific angular momentum as
\begin{eqnarray}
\tilde{E}^2 &=& c^2\left(1 - \frac{2\tilde{M}}{r} - \frac{8\pi \tilde{\Lambda}}{3}r^2\right)\left[\frac{\tilde{L}^2}{r^2} + c^2 \Delta^{-2/(2\omega+3)}\right], \\
\tilde{L}^2 &=& \frac{c^2 r^2}{(1-3\tilde{M}/r)}\left[\left(\frac{\tilde{M}}{r}-\frac{8\pi \tilde{\Lambda}}{3}r^2\right) - \frac{2}{(2\omega +3)}
\left(1 - \frac{\tilde{M}}{r} - \frac{16\pi \tilde{\Lambda}}{3}r^2\right)\right]\Delta^{-2/(2\omega+3)}.
\end{eqnarray}
Note that it is the rotation angular velocity (per unit test mass) $v_{\phi}$ rather than the specific angular momentum that we are after.
Thus consider, again on the equatorial plane, the {\it proper} rotation velocity given by
$v^2_{\phi} = \left(d\sigma_{3}/dt\right)^2$ with $ds^2 = g_{tt}dt^2 + d\sigma^2_{3}$ (i.e., $d\sigma^2_{3}$ denotes {\it spatial}
line element) and $dr = d\theta = 0$, namely
\begin{eqnarray}
v^2_{\phi} &=& g_{\phi \phi}\left(\frac{d\phi}{dt}\right)^2 = g_{\phi \phi}\frac{\left(d\phi/d\tau\right)^2}{\left(dt/d\tau\right)^2} \nonumber \\
&=& \Delta^{-2/(2\omega+3)}r^2 \frac{\tilde{L}^2/r^4}{(1 - 2\tilde{M}/r - 8\pi \tilde{\Lambda}r^2/3)^{-2}\tilde{E}^2/c^4}
\end{eqnarray}
where we used eqs.(32) and (33). Further, substituting eqs.(37) and (38) into eq.(39) yields
\begin{eqnarray}
v^2_{\phi} = &&c^2 \Delta^{-2/(2\omega+3)} \times \\
&&\frac{(1 - 2\tilde{M}/r - 8\pi \tilde{\Lambda}r^2/3)\left\{(\tilde{M}/r - 8\pi \tilde{\Lambda}r^2/3)-\frac{2}{(2\omega +3)}
(1 - \tilde{M}/r - 16\pi \tilde{\Lambda}r^2/3)\right\}}{(1 - 3\tilde{M}/r) + \left\{(\tilde{M}/r - 8\pi \tilde{\Lambda}r^2/3)-\frac{2}{(2\omega +3)}
(1 - \tilde{M}/r - 16\pi \tilde{\Lambda}r^2/3)\right\}}.  \nonumber
\end{eqnarray}
This is the fully relativistic expression for the rotation angular velocity (per unit test mass).
Lastly, in the outer region of BD scalar halo, where $r\sim 10 ~(kpc) \simeq 10^{23}~(cm) >> G_{0}M/c^2 \simeq 10^{16}~(cm)$ and hence
$\Delta = [r^2(1 - 8\pi \tilde{\Lambda} r^{2}/3) - 2\tilde{M}r]/r^2_{0} \simeq [r^2(1 - 8\pi \tilde{\Lambda} r^{2}/3)]/r^2_{0}$,  this fully relativistic expression for the rotation
angular velocity reduces to
\begin{eqnarray}
v^2_{\phi} = c^2 &&\left[\frac{(1 - 8\pi \tilde{\Lambda}r^2/3)}{(1 - 8\pi \tilde{\Lambda}r^2/3)-\frac{2}{(2\omega +3)}
(1 - 16\pi \tilde{\Lambda}r^2/3)}\right] \times \\
&&\left[\frac{8\pi \tilde{\Lambda}}{3}r^2 + \frac{2}{(2\omega +3)}
\left(1 - \frac{16\pi \tilde{\Lambda}}{3}r^2\right)\right]\left[\frac{r^2}{r^2_{0}}\left(1- \frac{8\pi \tilde{\Lambda}}{3}r^2\right)\right]^{-2/(2\omega+3)}.  \nonumber
\end{eqnarray}
Further, upon substituting a large-$\omega $ value, say, $\omega \sim 10^6$, one ends up with
\begin{eqnarray}
v^2_{\phi} &\simeq & c^2 \left[\frac{2}{(2\omega +3)} + \frac{(2\omega -1)}{(2\omega +3)}\frac{8\pi \tilde{\Lambda}}{3}r^2 \right]
\left\{\frac{r^2}{r^2_{0}}\left(1- \frac{8\pi \tilde{\Lambda}}{3}r^2\right)\right\}^{-2/(2\omega+3)} \nonumber \\
&=& \left[100 (km/s) \times \left\{\frac{r^2}{r^2_{0}}\left(1- \frac{8\pi G_{0}}{3c^4}\Lambda r^2\right)\right\}^{-(1/10^6)}  \right]^2 \\
&+& \left[c^2 \left(\frac{8\pi G_{0}}{3c^4}\Lambda r^2_{0}\right)\left(\frac{r}{r_{0}}\right)^2\right]\left\{\frac{r^2}{r^2_{0}}\left(1- \frac{8\pi G_{0}}{3c^4}\Lambda r^2\right)\right\}^{-(1/10^6)}
\nonumber
\end{eqnarray}
where as before the normalization factor $r_{0} \sim 10 ~(kpc) \simeq 10^{23} ~(cm)$ which represent a typical distance to the outer region of the dark halo has been introduced.
As has been discussed in the text, therefore, in order for this result to successfully describe the flattened rotation curve in the outer region of dark halo, $r \geq r_{0}$,
it should be imposed that
\begin{eqnarray}
\left(\frac{8\pi G_{0}\Lambda}{3c^4}r^2_{0}\right) << 10^{-6} \nonumber
\end{eqnarray}
which again yields $\Lambda << 1.45 \times 10^{-4} ~(erg/cm^3)$. Note that this is essentially the same result as what we have gotten using the Newtonian mechanics version of
the rotation velocity in section IV.
To summarize, even if we employ the fully relativistic expression for the rotation velocity, the conclusions we have reached in the previous section
essentially remain the same.

\subsection{Singularity along the Symmetry Axis: the Relativistic Bipolar Jet Interpretation}

Earlier in section II, we mentioned in advance that a rigorous study of the true nature of the singularity along the symmetry axis reveals the fact that
the bizzare singularity at $\theta = 0, ~\pi$ of the Schwarzschild-de Sitter-type solution in BD gravity theory can account for the relativistic
bipolar outflows (twin jets) extending from the central region of ``active galactic nuclei (AGNs)''. Thus in the present section, we shall demonstrate in an
explicit manner that this is indeed the case.  \\
In order to see if indeed the Schwarzschild-de Sitter-type spacetime solution in BD gravity theory allows for the occurrence of relativistic bipolar outflows,
we particularly explore the geodesic motion of a test particle in the immediate vicinity of the symmetry axis, $\theta = 0, ~\pi$.
Namely, consider the geodesic motion at $\theta \simeq \delta << 1$, (and hence $\sin \theta \simeq \sin \delta \equiv \epsilon << 1$).
Then the conserved quantities, i.e., the specific energy $\tilde{E}=E/m $ and the specific angular momentum $\tilde{L}=L/m $ of a test particle along
the geodesic are now given by
\begin{eqnarray}
\tilde{E} &=&
\left(\epsilon^2 \Delta \right)^{-2/(2\omega+3)}\left(1 - \frac{2\tilde{M}}{r} - \frac{8\pi \tilde{\Lambda}}{3}r^2\right)c^2\left(\frac{dt}{d\tau}\right) \gg  1, \\
\tilde{L} &=&
\left(\epsilon^2 \Delta \right)^{-2/(2\omega+3)}r^2 \epsilon^2 \left(\frac{d\phi }{d\tau}\right) = \left(\epsilon^2 \right)^{(2\omega+1)/(2\omega+3)}
\Delta^{-2/(2\omega+3)} r^2 \left(\frac{d\phi }{d\tau}\right) \ll  1. \nonumber
\end{eqnarray}
Namely, the specific energy of a test particle is extremely large and its specific angular momentum is very small there for generic $\omega$-value.
And it is rather obvious to see that mainly this can be attributed to the metric function factor $\left(\Delta \sin^{2}\theta \right)^{-2/(2\omega+3)}$
(where $\sin \theta$ is replaced by $\epsilon \ll 1$) which is responsible for the singular nature of the symmetry axis. Note first that the extremely small
specific angular momentum near the symmetry axis is rather expected (like in the Einstein gravity context) as it is basically due to the negligible arm length
from the symmetry axis. (Angular velocity $\left(d\phi /d\tau \right)$ there, however, may not be so small.) The extremely large specific energy near the
symmetry axis, however, is indeed something unexpected and thus is surprising as it does not happen in the Einstein gravity context
(i.e., the $\omega \to \infty$ limit) where the symmetry axis is perfectly regular.  Now, this study of the true nature of the singularity along the symmetry axis
leads us to suspect that perhaps the bizzare singularity at $\theta = 0, ~\pi$ of the Schwarzschild-de Sitter-type solution in BD gravity theory can account for the
relativistic {\it bipolar outflows (twin jets)} extending from the central region of active galactic nuclei (AGNs). That is, the curious singularity along the
symmetry axis seems harmless, after all. Rather, it turns out to be a pleasant surprise as it can explain a long-known puzzle in observed features of galaxies.
We now demonstrate that indeed this extremely large specific energy near the symmetry axis can be translated into the ultrarelativistic speed at which the test
particle moves along the symmetry axis. To this end, we start with $g_{\mu \nu}u^{\mu}u^{\nu} = -\kappa $ particularly near the poles, namely,
\begin{eqnarray}
-\kappa &=& \left(\epsilon^2 \Delta \right)^{-2/(2\omega+3)}
\left[- \left(1 - \frac{2\tilde{M}}{r} - \frac{8\pi \tilde{\Lambda}}{3}r^2\right)c^2 \left(\frac{dt}{d\tau}\right)^2 + r^2 \epsilon^2 \left(\frac{d\phi}{d\tau}\right)^2 \right] \nonumber \\
&+& \left(\epsilon^2 \Delta \right)^{2/(2\omega+3)}
\left[\left(1 - \frac{2\tilde{M}}{r} - \frac{8\pi \tilde{\Lambda}}{3}r^2\right)^{-1}\left(\frac{dr}{d\tau}\right)^2 + r^2 \left(\frac{d\delta}{d\tau}\right)^2 \right] \nonumber \\
&\simeq & -\left(\epsilon^2 \Delta \right)^{-2/(2\omega+3)}\left(1 - \frac{2\tilde{M}}{r} - \frac{8\pi \tilde{\Lambda}}{3}r^2\right)c^2 \left(\frac{dt}{d\tau}\right)^2  \\
&+& \left(\epsilon^2 \Delta \right)^{2/(2\omega+3)}\left(1 - \frac{2\tilde{M}}{r} - \frac{8\pi \tilde{\Lambda}}{3}r^2\right)^{-1}\left(\frac{dr}{d\tau}\right)^2 \nonumber
\end{eqnarray}
where we used $\delta << 1$ and $\epsilon << 1$. We are now ready to study the first integral of the timelike (for massive particles, $\kappa = c^2$) radial geodesic
equation given by
\begin{eqnarray}
\left(\frac{dr}{d\tau}\right)^2 &=&
\left(\epsilon^2 \Delta \right)^{-4/(2\omega+3)}\left(1 - \frac{2\tilde{M}}{r} - \frac{8\pi \tilde{\Lambda}}{3}r^2\right)c^2
\left[\left(1 - \frac{2\tilde{M}}{r} - \frac{8\pi \tilde{\Lambda}}{3}r^2\right)\left(\frac{dt}{d\tau}\right)^2 - \left(\epsilon^2 \Delta \right)^{2/(2\omega+3)}\right] \nonumber \\
&\simeq& \left(\epsilon^2 \Delta \right)^{-4/(2\omega+3)}\left(1 - \frac{2\tilde{M}}{r} - \frac{8\pi \tilde{\Lambda}}{3}r^2\right)^{2}c^2\left(\frac{dt}{d\tau}\right)^2.
\end{eqnarray}
Thus along the symmetry axis $\theta = 0, ~\pi$, the {\it proper} ejection velocity would be given by
\begin{eqnarray}
v^2_{r} &=& g_{rr}\left(\frac{dr}{dt}\right)^2 = g_{rr}\frac{\left(dr/d\tau\right)^2}{\left(dt/d\tau\right)^2}  \\
&=& c^2 \frac{1}{\left[\epsilon^2 (r/r_{0})^2\right]^{2/(2\omega +3)}}
\left(1 - \frac{2\tilde{M}}{r} - \frac{8\pi \tilde{\Lambda}}{3}r^2\right)^{(2\omega+1)/(2\omega+3)} \nonumber \\
&\simeq& c^2 \left(1 - \frac{2\tilde{M}}{r} - \frac{8\pi \tilde{\Lambda}}{3}r^2\right)   \nonumber
\end{eqnarray}
where we used eq.(45) and $\Delta = [r^2(1 - 8\pi \tilde{\Lambda} r^{2}/3) - 2\tilde{M}r]/r^2_{0}$ in the second line and $\omega \sim 10^6 \sim$ large in the last line.
Once again, it is interesting to note that if the value of the BD $\omega$-parameter were {\it not} large enough (like $\omega \sim 10^6$ which has been assigned
to reproduce the flattened rotation curve in the galactic halos), the ejection speed of test particles along the symmetry axis $\left(dr/dt \right)$ could encounter
the danger of exceeding the speed of light due to the factor $1/(\epsilon^2)^{2/(2\omega+3)}$ on the right hand side for $\left(dr/dt \right)$.
Lastly, for a typical galaxy with total mass $M\sim 10^{11}M_{\odot}$ and with the observed value of the cosmological constant $\Lambda \simeq 10^{-8} ~(erg/cm^3)$,
we have $2G_{0}M/c^2 \simeq 0.01 ~pc$ and $\left(8\pi G_{0}\Lambda /3c^4 \right)^{-1/2} \simeq 4 ~Gpc$ and hence we end up with
\begin{eqnarray}
v_{r} \simeq c \left(1 - \frac{2\tilde{M}}{r} - \frac{8\pi \tilde{\Lambda}}{3}r^2\right)^{1/2}
= c\left[1 - \frac{(0.01 ~pc)}{r} - \left(\frac{r}{4\times 10^9 ~pc}\right)^2 \right]^{1/2}.
\end{eqnarray}
Observationally, it is well-known that the typical size of the active galatic nuclei is less than 1 (pc) and the extent of typical galactic jets ranges from
several (kpc) to a few (Mpc). Thus the ejection speed of test particles along the symmetry axis (i) just outside the AGN, i.e., for $r \sim $ few (pc) is
$v_{r} \simeq c [1 - 0.01]^{1/2} \simeq 0.995 c$ and (ii) well-above the galactic plane, say, for $r \sim $ 10 (kpc) is
$v_{r} \simeq c [1 - 10^{-6} - 10^{-10}]^{1/2} \leq c$.
This completes the rigorous demonstration that the test particles move along the symmetry axis at nearly the speed of light. In addition, since the present analysis
of the geodesic motion is valid only in the immediate vicinity of the symmetry axis, it indicates the {\it collimation} of the outflow as well consistently with the
observation (see Fig.2).
\begin{figure}[hbt]
\centerline{\epsfig{file=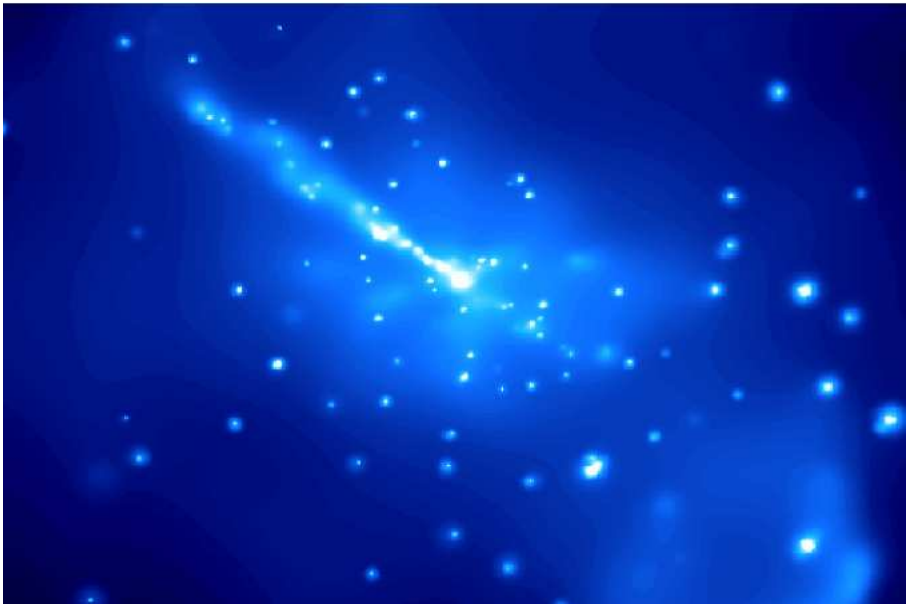, width=12cm, height=8cm}}
\caption{\ Centaurus A : An elliptical galaxy with an AGN and a
jet.}
\end{figure}

\section{Concluding remarks}

In the present work, the pure (i.e., with no ordinary matter) BD gravity with or without the cosmological constant $\Lambda $
has been demonstarted as a successful candidate for the theory of dark matter.
To summarize, the mysterious flattened rotation curves observed for so long in the outer region of
galactic halos, among others, have been the primary evidence for the existence of dark matter.
Therefore, it has been demonstrated in this work that
our model theory can successfully predict the emergence of dark matter halo-like configuration in terms of a self-gravitating static
and nearly spherically-symmetric spacetime solution to the BD field equations and reproduce the flattened rotation curve in the
outer region of this dark halo-like object in terms of the non-trivial energy density of the BD scalar field which was absent
in the context of general relativity where the Newton's constant is strictly a ``constant'' having no dynamics.
As stated earlier, there has been a consensus that unless one modifies either the general relativity, the standard theory of gravity,
or the standard model for particle physics, or both, one can never attain the satisfying understanding of the phenomena associated with
dark matter and dark energy. Therefore, our dark matter model presented in this work can be viewed as an attempt to modify the gravity
side alone in terms of the Brans-Dicke theory to achieve the goal. After all, the interesting lesson we learned from this study is the
fact that the simplest extension of general relativity which involves relaxing Newton's constant to a dynamical field
(i.e., the BD scalar field $\Phi (x)$) essentially makes all the differences.
Thus to conclude, from this success of ``BD scalar field as the dark matter'' to account
for the asymptotic flattening of galaxy rotation curves while forming galactic
dark matter halos {\it plus} the original spirit of the BD theory in which the BD scalar field is
prescribed {\it not} to have direct interaction with ordinary matter fields (in order not to interfere
with the great success of equivalence principle), we suggest that the Brans-Dicke theory of gravity
{\it is} a very promising theory of dark matter. And this implies, if we emphasize it once again, that dark matter
(and dark energy as well, see \cite{hongsu2}) might not be some kind of an unknown exotic ``matter'', but instead
the effect resulting from the space-time varying nature of the Newton's constant represented by the
BD scalar field. Even further, this successful account of the
phenomena associated with dark matter of the present universe via the BD gravity theory might be
an indication that the truly relevant theory of classical gravity at the present epoch is not
general relativity but its simplest extension, the Brans-Dicke theory with its generic parameter value
$\omega \sim 10^6$ fixed by the dark matter observation. This idea, however, should be taken with some caution as there are other phenomena
observed thus far (than the flattened rotation curves) which are suspected to be related to the effects of dark matter.
These include the galaxy/cluster lensing of distant quasars. Besides, our model theory for dark matter presented in this work is not without a flaw.
Earlier, from the rigorous study of the true nature of the singularity along the symmetry axis of this Schwarzschild-de Sitter-type spacetime, we suspected that
perhaps this  bizzare singularity at $\theta = 0, ~\pi$ can account for the relativistic bipolar outflows (twin jets) extending from the central region of
active galactic nuclei (AGNs). The relativistic bipolar outflows, however, have been observed only for some particular types of galaxies
such as AGNs and micro-quasars. That is, the jets do not seem to be a general feature of all types of galaxies.
Therefore the Schwarzschild-de Sitter-type solution in BD gravity theory employed in the present work appears to come with only a limited
descriptive power for the observed phenomena associated with galaxies.  At present in the absence of Birkhoff-type theorem guaranteeing the uniqueness of the
Schwarzschild-type solution in BD gravity theory, however, it seems worth looking for an alternative solution and proceeding with it toward the issue addressed in
the present work. \\
Therefore more extensive and careful study needs to be done to test the Brans-Dicke theory as a truly successful model theory
of dark matter and dark energy and we hope to report more along this line in the near future.

\vspace*{1cm}

\begin{center}
{\rm\bf Acknowledgements}
\end{center}

This work was supported in part by the Korea Research Council of Fundamental Science and Technology (KRCF),
Grant No. C-RESEARCH-2006-11-NIMS.

\vspace*{2cm}

\begin{center}
{\rm\bf Appendix : Halos of BD scalar field : the case of Einstein frame }
\end{center}

Note that in the present work, we ignore the presence of other types of ordinary matters.
Furthermore, we particularly consider the case when the cosmological constant term
(which is a dimensionful parameter in the action) is neglected as well.
Namely, we are dealing with the pure Brans-Dicke theory of gravity and we particularly
restrict our interest to the original spirit of Brans and Dicke according to which the BD scalar
field $\Phi $ is prescribed to remain strictly massless without any self-interaction potential.
Then in such a context, the study of possible role of BD scalar field as a dark matter component in
Einstein frame, which is related to the Jordan frame (which we have been working in) by the Weyl
rescaling (i.e., conformal transformation plus field redefinition), would be relevant as well.
Thus we shall turn to this issue in this appendix. \\
We now begin with the Weyl-rescaling given by
\begin{eqnarray}
g_{\mu\nu} &=& \Omega^2(x)\tilde{g}_{\mu\nu}, ~~~\Phi = M^2_{pl}e^{\Psi/\Psi_{0}} \\
{\rm with} ~~~\Omega^2(x) &=& \frac{M^2_{pl}}{\Phi}, ~~~\Psi^2_{0} = (2\omega+3)
\nonumber
\end{eqnarray}
where $M^{2}_{pl}=1/G_{0}$ denotes the Planck mass.
Under this Weyl-rescaling, the action in the Jordan frame given earlier in eq.(1) but without the
$\Lambda $-term transforms to
\begin{eqnarray}
\tilde{S} = \int d^4x \sqrt{\tilde{g}}{M^2_{pl}\over 16\pi}\left[\tilde{R} -
\frac{1}{2} \tilde{g}^{\mu\nu}{\partial_{\mu}\Psi \partial_{\nu}\Psi }\right].
\end{eqnarray}
Then extremizing this action with respect to the metric $\tilde{g}_{\mu \nu}$ and the redefined
BD scalar field $\Psi $ yields the classical field equations given respectively by
\begin{eqnarray}
\tilde{G}_{\mu \nu} &=& \tilde{R}_{\mu \nu} - {1\over 2}\tilde{g}_{\mu \nu}\tilde{R} =
8\pi \tilde{T}^{BD}_{\mu \nu},
~~~\tilde{g}^{\alpha \beta}\nabla_{\alpha}\nabla_{\beta}\Psi = 0  ~~~{\rm where}  \nonumber \\
\tilde{T}^{BD}_{\mu \nu} &=& {1\over 16\pi}\left[\nabla_{\mu}\Psi \nabla_{\nu}\Psi
- {1\over 2}\tilde{g}_{\mu \nu}\left(\tilde{g}^{\alpha \beta}\nabla_{\alpha}\Psi \nabla_{\beta}\Psi \right)\right].
\end{eqnarray}
The action and the classical field equations in this Einstein frame appear to take the forms of those of massless
scalar field theory coupled minimally to Einstein gravity. A caution, however, needs to be exercised.
That is, despite how it seems, this theory in the Einstein frame is really the (pure) BD gravity in disguise.
Indeed, close inspection of eq.(48) above reveals that the Weyl rescaling becomes trivial only for
$\omega \rightarrow \infty$, which is the Einstein gravity limit (with no scalar field), but for finite $\omega$
values this theory is just the conformal rescaling of the (pure) BD gravity.
As such, the solution to these field equations in the Einstein frame in eq.(50) should not be looked for independently
without referring to that in the Jordan frame. Indeed, it would simply be the {\it same} Weyl rescaling of the solution
constructed in the Jordan frame given in eq.(14), namely,
\begin{eqnarray}
d\tilde{s}^2 &=& \tilde{g}_{\mu \nu}dx^{\mu}dx^{\nu}
= \left(\frac{\Phi}{M^2_{pl}}\right)g_{\mu \nu}dx^{\mu}dx^{\nu} \nonumber\\
&=& \left[-\left(1 - {2M\over r}\right)dt^2 + r^2 \sin^2 \theta d\phi^2 \right]
+ \left(\Delta \sin^{2}\theta \right)^{4/(2\omega+3)}
\left[\left(1 - {2M\over r}\right)^{-1}dr^2 + r^2 d\theta^2 \right], \nonumber \\
\Psi (r, \theta) &=& \Psi_{0}\ln {\left(\Phi/M^2_{pl}\right)} \\
&=& 2(2\omega+3)^{-1/2}\ln {(\Delta \sin^{2}\theta)}.  \nonumber
\end{eqnarray}
We now attempt to address the same issues as we did earlier while working in the Jordan frame, i.e.,
whether the BD scalar field can cluster into dark matter halo-like objects and reproduce flattened
rotation curves. Firstly, once again it is obvious that the BD scalar can cluster into
halo-like configuration as it can be represented by this Brans-Dicke-Schwarzschild solution in the
Einstein frame given in eq.(51). Secondly, we turn to the behavior of the rotation curve in the outer region of
this halo-like configuration. To this end, again we restore both the Newton's constant $G_{0}$ and the
speed of light $c$ to work in the CGS unit as before by multiplying the factor $(c^4/G_{0})$ to the
BD scalar field energy-momentum tensor in eq.(50) and by replacing $M\rightarrow G_{0}M/c^2 \equiv \tilde{M}$
in the BDS solution in Einstein frame given in eq.(51). Next, by substituting the BDS solution in eq.(51)
into the BD scalar energy-momentumm tensor in eq.(50) and then setting it equal to (15), again we can read off
the energy density and the pressure components of the BD scalar field imperfect fluid as
\begin{eqnarray}
\rho &=& {c^2\over 2\pi G_{0}}{1 \over (2\omega + 3)}
\left(\Delta \sin^2 \theta \right)^{-4/(2\omega+3)}
\left[\frac{(r-\tilde{M})^2}{r^2 \tilde{\Delta} } + \frac{\cos^2 \theta }{r^2 \sin^2 \theta }\right],  \nonumber \\
P_{r} &=& {c^4\over 2\pi G_{0}}{1 \over (2\omega + 3)}
\left(\Delta \sin^2 \theta \right)^{-4/(2\omega+3)}
\left[\frac{(r-\tilde{M})^2}{r^2 \tilde{\Delta} } - \frac{\cos^2 \theta }{r^2 \sin^2 \theta }\right], \\
P_{\theta} &=& -P_{r}, ~~~P_{\phi} = -c^2 \rho, \nonumber \\
T^{r}_{\theta} &=& \Delta T^{\theta}_{r} =
{c^4\over \pi G_{0}}{1 \over (2\omega + 3)}
\left(\Delta \sin^2 \theta \right)^{-4/(2\omega+3)}
\frac{(r-\tilde{M})}{r^2}\cot \theta . \nonumber
\end{eqnarray}
Now, in order to study the behavior of the rotation curves in the outer region of dark halos of
typical galaxies, we approximate the expression for the energy density of the BD scalar field
at large-$r$, say, $r >> G_{0}M/c^2$ and it is
$\rho \simeq c^2 (2\pi G_{0}(2\omega + 3))^{-1}(r^2\sin^2 \theta)^{-1}
\left(\Delta \sin^2 \theta \right)^{-4/(2\omega+3)}$.
Then the mass function is given by
$M(r) = \int^{2\pi}_{0}d\phi \int^{\pi -\epsilon}_{\epsilon}d\theta \int^{r}_{0}dr
\sqrt{\tilde{g}_{rr}\tilde{g}_{\theta \theta}\tilde{g}_{\phi \phi}}\rho (r, \theta) =
\left(c^2/G_{0}(2\omega +3)\right)f(\delta)r$. Note here that we already have $M(r)\sim r$
regardless of the value of BD $\omega$-parameter in the present case of Einstein frame and hence
\begin{eqnarray}
v^2(r) = \frac{G_{0}M(r)}{r} = c^2 \frac{1}{(2\omega +3)}f(\delta)
\end{eqnarray}
where this time $f(\delta) \equiv \int^{\pi -\epsilon}_{\epsilon}d\theta \sin^{-1}\theta
= 2\int^{1-\delta}_{0}dx [1-x^2]^{-1} \simeq O(1)$ with $\epsilon, ~\delta <<1$.
This is already a flattened rotation curve, and if we further take, say, $\omega \sim 10^6$
once again, we get
\begin{eqnarray}
v(r) \simeq 100 (km/s).
\end{eqnarray}
Therefore, regardless of whether we work in the Jordan or in the Einstein frame, the BD scalar
field imperfect fluid always reproduces the flattened rotation curve. And from this observation, the
BD scalar field appears to play the role of dark matter component in the galactic halos. \\
Consider next, the equation of state of this BD scalar field imperfect fluid in this Einstein frame
\begin{eqnarray}
w = {P\over c^2\rho }
= \frac{[(r-\tilde{M})^2/r^2 \tilde{\Delta} ] - [\cos^2 \theta /r^2 \sin^2 \theta ]}
{[(r-\tilde{M})^2/r^2 \tilde{\Delta} ] + [\cos^2 \theta /r^2 \sin^2 \theta ]}
\end{eqnarray}
where $P = P_{r}$. Just like in the earlier study in the Jordan frame, $P = w(r, \theta)c^2 \rho $ with
$w(r, \theta) \sim O(1)$ meaning that this BD scalar fluid is still a barotropic
fluid but with position-dependent coefficient $w(r, \theta)$. And once again, although the BD scalar
field is a candidate for dark matter, this equation of state indicates that in general, particularly in the
vicinity of the individual galaxy, it is not quite a dust.
In the outer halo region, i.e., at large-$r$, however, $w\simeq 0$ (as $P\simeq 0$ in the numerator) meaning that
indeed the BD scalar fluid in the Einstein frame exhibits the property of {\it dust} matter everywhere including
on the galactic (equatorial) plane regardless of the value of BD parameter $\omega $. \\
We now compare the results obtained earlier in the Jordan frame with those here in the Einstein frame. Firstly, in both frames,
the BD scalar field imperfect fluid appears to play the role of dark matter component in a successful manner
as it always reproduces the flattened rotation curve in the outer region of galactic halos. Secondly, the
equations of state of the BD scalar fluid in both frames indicate that it is generally not quite a dust and this point
implies that the BD scalar field appears to be an ``exotic'' type of dark matter possessing non-standard equation
of state generally in the vicinity of individual galaxy.
In the outer halo region, i.e., at large-$r$, however, it is interesting to note that
$w \simeq 0$, namely a dust matter for large-$\omega$ parameter value and particularly on the galactic (equatorial) plane, where
most of the gases or stars orbit around the galaxy center, in the case of Jordan frame whereas it is $w\simeq 0$
everywhere, {\it not} just on the galactic (equatorial) plane, and {\it regardless} of the value of $\omega $
in the case of Einstein frame. \\
One might wonder how this somewhat delicate discrepancy in the nature of equation of state for the BD scalar fluid in the
two conformal frames arises. We now attempt to provide one possible cause that may lead to this slight discrepancy.
Indeed, close inspection reveals that
it can be attributed to the non-standard form of the BD scalar field energy-momentum tensor in the Jordan frame.
That is, in the expression for the BD scalar field energy-momentum tensor given in eq.(2), we can realize that it
involves both the terms quadratic in first derivatives of the BD scalar field $\Phi$ coming from its kinetic energy term
in the action and those linear in second derivatives of $\Phi$ coming from its non-minimal coupling term to curvature,
$\sim \sqrt{g}\Phi R$. (Note that its standard expression for the ordinary scalar field (like that in the Einstein frame given
in eq.(50) for the Weyl rescaled BD scalar field $\Psi$) involves only the terms quadratic in first derivatives of the
scalar field.) And these {\it anomalous} terms linear in second (covariant) derivatives of $\Phi$ seem to be the ones that
make the difference as they turn out to render the pressure of the BD scalar imperfect fluid {\it negative} at
large-$r$ and particularly for large-$\omega$ while {\it zero} on galactic equatorial plane in the Jordan frame.
In both frames, however, the BD scalar field imperfect fluid appears to reproduce generic features of the dark matter in the
outer region of dark halo of the individual galaxy.  \\
Note that in the other case where the cosmological constant is present, the issue of selecting from Jordan or Einstein frame that we
just discussed becomes irrelevant to address. And it is because when a dimensionful parameter such as the cosmological
constant (with mass dimension 4 in the Planck unit) is present in the theory action, the Weyl-rescaling of the action/field equations
to go from the original Jordan frame, say, to Einstein frame does not leave the physics invariant. Therefore in section IV in the text,
we just worked in the original Jordan frame.


\noindent

\begin{center}
{\rm\bf References}
\end{center}


\begin{thebibliography}{99}

\bibitem{bd} C. Brans and C. H. Dicke, Phys. Rev. {\bf 124}, 925 (1961).

\bibitem{will} C. M. Will, {\it Was Einstein Right ?},
                   (Basic Books, Inc., Publishers/New York, 1986).

\bibitem{weinberg} S. Weinberg, {\it Gravitation and Cosmology}, (John Wiley and Sons, Inc., 1972).

\bibitem{hongsu2} H. Kim, Phys. Lett. {\bf B606}, 223 (2005) ;
                  Mon. Not. Roy. Astron. Soc. {\bf 364}, 813 (2005).

\bibitem{tina} R. N. Tiwari and B. K. Nayak, Phys. Rev. {\bf D14}, 2502 (1976) ;
               J. Math. Phys. {\bf 18}, 289 (1977).

\bibitem{sinrai} T. Singh and L. N. Rai, Gen. Rel. Grav. {\bf 11}, 37 (1979).

\bibitem{mm} R. A. Matzner and C. W. Misner, Phys. Rev. {\bf 154}, 1229 (1967).

\bibitem{mispan} R. M. Misra and D. B. Pandey, J. Math. Phys. {\bf 13}, 1538 (1972).

\bibitem{hongsu1} H. Kim, Phys. Rev. {\bf D60}, 024001 (1999).

\bibitem{hongsu3} H. Kim and H. M. Lee, Int. J. Mod. Phys. A20, 6461 (2005).

\bibitem{stress} F. E. Schunck and E. W. Mielke, Class. Quantum Grav. {\bf 20}, R301 (2003).

\bibitem{rc} See for instance, over 100 rotation curve fits given in
             J. R. Brownstein and J. W. Moffat, Astrophys. J. {\bf 636}, 721 (2006) ;
             see also, V. Sahni, Lect. Notes Phys. 653, 141 (2004).

\bibitem{halo1} S. McGaugh, V. Rubin, and E. de Block, Astron. J. {\bf 122}, 2381 (2001).

\bibitem{halo2} M. Persic, P. Salucci, and F. Stel, Mon. Not. Roy. Astron. Soc. {\bf 281}, 27 (1996) ;
                E. Corbelli and P. Salucci, astro-ph/9909252 ;
                Y. Sofue and V. Rubin, Ann. Rev. Astron. Astrophys. 39, 137 (2001).

\bibitem{tf} R. B. Tully and J. R. Fisher, Astron. Astrophys. {\bf 54}, 661 (1977).

\bibitem{khalo} C. Armendariz-Picon and E. A. Lim, JCAP 0508, 007 (2005) ; F. E. Schunk, astro-ph/9802258.

\bibitem{wmap} C. L. Bennett et al., Astrophys. J. Suppl. {\bf 148}, 1 (2003) ;
               G. Hinshaw et al., Astrophys. J. Suppl. {\bf 148}, 135 (2003) ;
               D. N. Spergel et al., Astrophys. J. Suppl. {\bf 148}, 175 (2003).

\bibitem{lambda} E. J. Copeland, M. Sami, and S. Tsujikawa, hep-th/0603057.

\bibitem{bpt} J. M. Bardeen, W. H. Press, and S. A. Teukolsky, Astrophys. J. {\bf 178}, 347 (1972)

\end{thebibliography}
\end{document}